\documentclass[onecolumn]{IEEEtran}
\usepackage{graphicx,subfig,amsthm,amsmath,latexsym,amssymb,times,bm}
\usepackage{float,epsfig,multirow,rotating,times,verbatim,wrapfig,booktabs}
\usepackage{color,xr,array,algpseudocode,setspace}
\definecolor{menucolor}{rgb}{0.1,0.52,0.47}
\definecolor{urlcolor}{rgb}{0.85,0.37,0.01}
\definecolor{runcolor}{rgb}{0.46,0.44,0.701}
\definecolor{filecolor}{rgb}{0.2,0.5,0.01}
\definecolor{linkcolor}{rgb}{0.12,0.47,0.70}
\definecolor{citecolor}{rgb}{0.55,0.36,0.01}
\definecolor{anchorcolor}{rgb}{0.4,0.4,0.4}
\usepackage[colorlinks=true, urlcolor=urlcolor, linkcolor=linkcolor,citecolor=citecolor,filecolor=filecolor, anchorcolor=anchorcolor,menucolor=menucolor]{hyperref}
\usepackage{nameref}
\usepackage{zref-xr}
\usepackage[nocompress]{cite}
\usepackage{cite}
\zxrsetup{
  tozreflabel=false,
  toltxlabel=true,  
}

\newcommand{\citep}{\cite}
\newcommand{\citet}{\cite}

\DeclareMathOperator{\E}{E}

\DeclareMathOperator{\diag}{diag}

\DeclareMathOperator{\trace}{tr}

\newcommand{\I}{\mathbb{I}}
\newcommand\given[1][]{\:#1\vert\:}

\newcommand{\Norm}{\mathcal{N}}
\newcommand{\Gam}{\text{Gamma}}

\newcommand{\R}{\mathbb{R}}

\begin{document}

\title{Fast Model-based Clustering of Partial Records}%
\author{Emily~M.~Goren and Ranjan~Maitra%<-this % stops a space 
  \thanks{E.~M.~Goren and
    R. Maitra are with the Department of Statistics, Iowa State
    University, Ames, 50011-1090, Iowa, USA. Email: \{egoren,maitra\}@iastate.edu.}
    \thanks{}

%\thanks{\copyright 201x IEEE. Personal use of this material is
%permitted. However, permission to use this material for any other
%purposes must be obtained from the IEEE by sending a request to
%pubs-permissions@ieee.org.} \thanks{Manuscript received xxxx xx,201x;
%revised xxxxxxxx xx, 201x.   First published xxxxxxxx x, xxxx,
%current version published   yyyyyyyy y, yyyy} \thanks{}  \thanks{Digital Object Identifier}
}
%\markboth{%IEEE Transactions on Knowledge and Data
          %Engineering,~Vol.~?, No.~?, January~201?}
%}%
%{Goren: Robust Efficient Model-based Clustering  Partial Records}

%--------------------------------------- Abstract --------------------------

\IEEEcompsoctitleabstractindextext{%

\begin{abstract}
Partially recorded data   are frequently
  encountered in many applications and usually clustered by first removing
  incomplete cases or features 
  with missing values, or by imputing missing values, followed by
  application of a clustering algorithm to the resulting altered dataset. Here, we develop clustering methodology through a model-based
  approach using the  marginal density for the observed values,
  assuming 
  a finite mixture model of multivariate $t$ distributions. We compare
  our approximate algorithm to the corresponding full expectation-maximization
  (EM) approach that considers the missing values in the incomplete
  dataset and makes a missing at random (MAR) assumption, as well as
  case deletion and imputation methods. Since only the observed values are
  utilized, our approach is computationally more efficient than
  imputation or full EM. Simulation studies demonstrate that our
  approach has favorable recovery of the true cluster partition
  compared to case deletion and imputation under various missingness
  mechanisms, and is at least competitive with the
  full EM approach, even when MAR assumptions are violated.
  Our methodology is demonstrated on a problem of
  clustering gamma-ray bursts and is implemented at \url{https://github.com/emilygoren/MixtClust}.
\end{abstract}

}
\maketitle

\IEEEdisplaynotcompsoctitleabstractindextext
%\doublespacing

\section{Introduction}\label{sec:intro}
	Cluster analysis partitions data into groups 
        % ``clusters''
        of similar observations in an unsupervised manner
        and commonly without knowledge of the total number of
        clusters. Clustering applications appear in many fields,
        including medical imaging \citep{maitra01}, gene expression
        \citep{witten11}, microbiome studies \citep{turnbaugh07},
        crime analysis \citep{agarwal13}, and astronomy
        \citep{feigelsonandbabu98}. Prominent approaches to cluster
        analysis can be grouped into centroid-based methods such as
        $k$-means \citep{macqueen67}, hierarchical clustering
        \citep{ward63}, and model-based methods
        \citep{melnykovandmaitra10}.  %Clustering is regarded as an
                                %instance of unsupervised learning,
                                %since the cluster labels         are
                                %unknown: however,  semi-supervised
                                %extensions         when some label
                                %information is available have also
                                %been developed \citep{basu04,           chapelle2009}.
        Refer to \citet{hartigan75,kettenring06,xuandwunsch09,melnykovandmaitra10,mcnicholas16,bouveryonetal19}
        for a comprehensive introduction to the rich topic of
        clustering.
%        , we refer the reader to \citet{hartigan75,kettenring06} and \citet{xuandwunsch09}.

	In practice, real datasets may have missing values or
        otherwise partially observed records that complicate
        the validity and application of standard statistical
        methodology. Missingness may result from diverse causes, with
        an underlying mechanism of one of three types: missing
        completely at random (MCAR), missing at random (MAR), or not
        missing at random (NMAR) \citep{rubin76}. Under MCAR, the
        probability that a case (record, sample, or observation) is
        missing feature (variable, attribute, dimension) values does
        not depend on either the observed or missing feature
        values. The mechanism is MAR when this probability depends on
        the observed feature values, but \emph{not} the missing
        feature values, and NMAR when the probability depends on \emph{both} observed and missing feature values. Notably, MCAR data are also MAR; if the data are not MAR, then they are NMAR. Strategies for analysis of data with missing values are often critically dependent on the missingness mechanism, and clustering is no exception.
	
	For clustering problems, the most common (and often expedient)
        treatment of missing values is deletion, on either a case or
        feature basis, or imputation \citep{little14,
          wagstaff05}. Given a dataset with $n$ cases and $p$
        features, case deletion removes all cases with any missing
        values across the $p$ features, leading to a reduced dataset
        with $n'<n$ cases that are fully observed for all the $p$
        original features. After a clustering algorithm has been
        applied to the resulting reduced dataset of complete cases,
        the remaining $n - n'$ incomplete cases can be assigned to the
        obtained cluster partition, for example, by using a partial
        distance \citep{hathawayandbezdek01} or marginal posterior
        probability \citep{chattopadhyayandmaitra18} approach. An
        alternative deletion approach is executed on a feature-wise
        basis by discarding any features that are not observed for all
        $n$ cases, resulting in a dataset of $n$ cases but $p'<p$
        features, on which a clustering algorithm can be applied to
        directly cluster all cases \citep{chattopadhyay17}. While
        attractive for their ease of implementation, both data
        exclusion schemes make an assumption of a MCAR mechanism,
        violation of which leads to reduced clustering
        performance. Even if data are MCAR, deletion may yield poor
        clustering performance due to loss of information. In
        contrast, imputation approaches \citep{buuren11, donders06,
          honaker11, park16} for clustering replace each missing value
        with a predicted value to produce a completed dataset that
        can be supplied to the desired algorithm to cluster all $n$
        cases. Critically, this approach treats the imputed values as if they were observed values, and thus ignores any error and uncertainty associated with the fact that they are not the actual values. Obtaining a suitable method for imputation can be difficult because the most appropriate choice likely depends on the unknown cluster partition. As a consequence, use of imputation has been shown to substantially diminish clustering performance \citep{wagstaff05}.
	
	The drawbacks of deletion and imputation have prompted
        clustering approaches that incorporate the
        missing data structure, yet       use all the observed entries
        in the dataset. \citet{dixon79} and  \citet{zhang14} used a partial
        distance to measure the distance between differentially
        observed cases. \citet{sarkarandleong01} extended the fuzzy
        $k$-means algorithm for incomplete cases but after
        imposing soft constraints based on estimating the distance
        between incomplete cases and cluster centers. % in an
                                % extension of the fuzzy
                                % $k$-means algorithm.
%        A weighted $c$-means algorithm using partial distance was
%        proposed by \citet{zhang14}.
        In fuzzy clustering,
        \citet{siminski2013clustering, simnski2014rough,
          siminski2015rough} utilized complete cases to define cluster
        centers and weights and multiply imputed the incomplete cases
        in the objective function. The $k$-means extension of
        \citet{wagstaff04} also used soft constraints defined by the
        partially observed features, but this requires at least one
        feature to be observed across all cases.  The $k$-POD
        algorithm \citep{chi2016k} employs majorization-minimization
        \citep{lange16} to minimize the objective function of
        $k$-means using partial distances for incomplete
        cases. Recently, \citet{lithio18} developed the $k_m$-means
        algorithm, which generalized the $k$-means algorithm of
        \citet{hartiganandwong79} to include partially-observed records. These
        approaches all utilized the Euclidean metric for measuring the
        distance between observations and cluster centers, in the
        process assuming hyperspherical-shaped clusters and giving up on robustness to
        outliers.  
	
	Model-based clustering using finite mixtures of the multivariate
        Gaussian \citep{banfieldandraftery93, celeuxandgovaert95} or
        $t$ distributions \citep{mclachlan98, peelandmclachlan00}
        allow for hyperellipsoidal-shaped clusters \citep{lindsay95}
        through use of the Mahalanobis distance \citep{mahalanobis36}
        and have a long history of successful application. Compared to
        the Gaussian distribution, the $t$ distribution confers
        greater resistance to outliers through its wider tails, and is
        therefore often the {\em de facto} choice for model-based
        clustering. For clustering of incomplete data,
        \citet{wangetal04} proposed finite mixture modeling using
        multivariate $t$ distributions and designed an
        expectation-maximization (EM) algorithm \citep{dempsteretal77}
        for both estimating mixture model parameters and treatment
        of the missing values. 
        \citet{lin2014} extended this work to incorporate the
        eigen-decomposed covariance        structure of \citet{banfieldandraftery93}. To
        better fit asymmetrical-shaped clusters, \citet{wang15}
        developed an approach using skew-$t$ distributions while
        \citet{wei19} extended this work with a eigen-decomposed
        covariance structure for skew-$t$-and generalized hyperbolic
        distributions. Importantly, all of these works included the
        missing values in the formal incomplete dataset within an EM
        algorithm. This may be computationally burdensome and lack
        robustness to the MAR assumption since, in the expectation
        step (E-step), the observed feature values inform those that are missing. When the data are NMAR, the observed values are not directly informative of those that are missing because their values are related to their own missingness.
		
	In this paper, we~(Section~\ref{sec:methods}) propose model-based clustering of partially
        recorded data using finite  multivariate-$t$ mixture
        distributions for only the observed values by integrating out the missing values
        and excluding them from our clustering algorithm
        calculations. An alternating expectation-conditional
        maximization (AECM) algorithm \citep{mengandvandyk97} 
        implements our approach %for general covariance structures and
        and is seen  %         advantages of our approach are that it
        in simulation studies~(Section~\ref{sec:sim}) to         
        reduce computational complexity when
        compared to a full EM approach and also to %be more robust
        % to severe
        do well under violations of the MAR mechanism. %We detail
                                %our approach in
                                %Section~\ref{sec:methods}.
        %Simulation studies in        Section~\ref{sec:sim} assess
        %computational speed and performance of our         method
        %under different missingness          mechanisms.
        Section~\ref{sec:app} uses our methodology to cluster
        gamma-ray bursts data. We conclude with some discussion in
        Section~\ref{sec:discussion}. Two appendices have additional  details.

\section{Methodology}\label{sec:methods}

\subsection{Background and Preliminaries}\label{sec:model}
We begin by introducing our model and relevant notation for the
problem of clustering the dataset $\bm{y} \in \R^{n \times p}$
consisting of $n$ cases and $p$ features into $K$ clusters, allowing
for missing values in $\bm{y}$. For now, we treat $K$ as known,
postponing discussion on choosing $K$ to later. Assume that the cases are independent and arise from a finite mixture of $t$ distributions described by the density
%\begin{equation} \label{eq:mod}
$f(\bm{y}_i \given \bm{\Theta}) = \sum_{k=1}^K \pi_k t_p(\bm{y}_i;
~\bm{\mu}_k, \bm{\Sigma}_k, \nu_k), \quad i = 1, \dots, n$
%\end{equation}
where $\pi_k \in (0,1)$ is the proportion represented by the $k^{th}$
cluster with $\sum_{k=1}^K\pi_k = 1$, and $t_p(\cdot; \bm{\mu}, \bm{\Sigma}, \nu)$ is the $p$-variate $t$-density with mean $\bm{\mu} \in \R^p$, positive-definite real $p \times p$ dispersion matrix $\bm{\Sigma}$, and degrees of freedom $\nu > 0$ defined as
\begin{equation}\label{eq:mvt}
t_p(\bm{y}; \bm{\mu}, \bm{\Sigma}, \nu) 
= \frac{\Gamma\left(\frac{\nu+p}{2} \right)} {\Gamma\left(\frac{\nu}{2}\right) \nu^{\frac{p}{2}}\pi^{\frac{p}{2}} \lvert \bm{\Sigma}\rvert^{\frac{1}{2}}} \left[ 1 + \frac{1}{\nu} \Delta(\bm{y}, \bm{\mu}, \bm{\Sigma}) \right]^{-\frac{\nu+p}{2}}, \quad \bm{y} \in \R^p
\end{equation}
where $\Delta(\cdot)$ is the Mahalanobis distance \citep{mahalanobis36}  given by
$%\begin{equation}\label{eq:maha}
\Delta(\bm{y}, \bm{\mu}, \bm{\Sigma}) = (\bm{y} - \bm{\mu})' \bm{\Sigma}^{-1}(\bm{y} - \bm{\mu}).
$%\end{equation}
%To ensure that \eqref{eq:mod} is a valid likelihood, we restrict $\sum_{k=1}^K\pi_k = 1$.

In missing data problems, the $p$-dimensional records are only
partially observed and we %seek to
leverage the observed values for
clustering. WLOG, suppose each $p$-vector is
decomposed into observed and missing components as $\bm{y}_i =
(\bm{y}_i^o, \bm{y}_i^m),$ where $\bm{y}_i^o \in \R^{p_i^o}$ is the
observed component and $\bm{y}_i^m \in \R^{p - p_i^o}$ is the missing
component for each observation $i = 1, \dots, n$. Define observed and
missing component selection matrices $\bm{O}_i$ and $\bm{M}_i$,
respectively, such that $\bm{O}_i$  extracts the observed component
from $\bm{y}_i$ and has dimension $p_i^o \times p$, and $\bm{M}_i$, of
dimension $(p-p_i^o) \times p$, extracts the missing component from
$\bm{y}_i$. Then 
$%\begin{equation*}%\label{eq:selmat}
\bm{y}_i^o = \bm{O}_i\bm{y}_i, \quad \bm{y}_i^m = \bm{M}_i\bm{y}_i.
$%\end{equation*}
These two matrices together account for the entire
$p$-dimensional vector through the property $\bm{O}_i'\bm{O}_i +
\bm{M}_i'\bm{M}_i = \bm{I}_p$. Omitting the vacuous case where no
features are observed, there are $\sum_{l=0}^{p-1} {p \choose l} =
2^p-1$ unique patterns of missingness possible for each case. 
The marginal density~\citep{lin2009} of the
observed values in the $i$th observation record is 
\begin{equation}\label{eq:obsmod}
f\left(\bm{y}_i^o \given \bm{\Theta}\right) 
= \sum_{k=1}^K \pi_k t_{p_i^o}\left(\bm{y}_i^o;~ \bm{\mu}_{ik}^o, \bm{\Sigma}_{ik}^{oo}, \nu_k\right), \quad i = 1, \dots, n
\end{equation}
where $ \bm{\mu}_{ik}^o = \bm{O}_i\bm{\mu}_k$ and
$\bm{\Sigma}_{ik}^{oo} = \bm{O}_i\bm{\Sigma}_k\bm{O}_i'$ are the
mean and dispersion of the observed components of the $i$th record in
the $k^\text{th}$ cluster.
\paragraph*{Comment} Our development above uses $\bm{O}_i$ as
a non-stochastic selection matrix to specify the presence of  measurement for the $i$th record, and as a computational
device. As pointed out by a reviewer, a more accurate development would
involve incorporating the stochastic nature of $\bm{O}_i$ to capture
the missingness mechanism if such were 
available. However, incorporating such a mechanism is
application-specific, and so for this paper, we have assumed that the
missingness mechanism is unavailable to us, and inherently
 MCAR.

 \subsection{EM Algorithm for Parameter Estimation}
 \subsubsection{Complete data and log likelihood} \label{sec:emcomplete}
A EM algorithm for parameter estimation can be formulated using \eqref{eq:obsmod} by specifying a so-called complete dataset and corresponding log likelihood. Since missing values are omitted in \eqref{eq:obsmod}, the  ``actual'' missing values $\{\bm{y}_i^m:i = 1,\dots,n\}$ are not part of the complete dataset; instead we include the ``conceptual'' missing values of class memberships and characteristic weights that we now introduce. Finite mixture modeling approaches to clustering can be recast as a problem of missing cluster membership labels. To this end, we define the latent class membership indicators
\begin{equation}\label{eq:latentids}
z_{ik} = \I(\text{case $i$ belongs to class $k$}), \quad i=1,\dots,n;~ k = 1, \dots, K
\end{equation}
where $\I(\cdot)$ denotes the indicator function. As an unsupervised learning task, all the $z_{ik}$'s are missing. To devise an EM algorithm, we also utilize the multivariate Gaussian-gamma mixture formulation of the multivariate $t$ distribution~\citep{cornish54}, rewriting \eqref{eq:mvt} as
\begin{equation}\label{eq:normalgamma}
t_p(\bm{y}; \bm{\mu}, \bm{\Sigma}, \nu) 
= \int_0^\infty \phi(\bm{y}; \bm{\mu}, \bm{\Sigma}/w) q(w; \nu/2, \nu/2)~\text{d}w, \quad \bm{y} \in \R^p
\end{equation}
where
$%\begin{equation*}\label{eq:normal}
\phi(\bm{y}; \bm{\mu}, \bm{\Sigma}/w) 
= { \lvert \bm{\Sigma}\rvert^{-\frac{1}{2}} \left(
  2\pi/w\right)^{-p/2} }\exp\left( - \frac{w}{2}  \Delta(\bm{y},
\bm{\mu}, \bm{\Sigma})  \right),   \mbox{ for }\bm{y}\in \R^p,
$%\end{equation*}
 is the $p$-variate Gaussian distribution with mean $\bm{\mu}$ and variance $\bm{\Sigma}/w$ denoted by $\mathcal{N}_p\left( \bm{\mu}, \bm{\Sigma}/w \right)$, and
 $%\begin{equation*}\label{eq:gamma}
q(w; \nu/2, \nu/2) 
= \{(\nu/2)^{\nu/2}/\Gamma(\nu/2)\} \exp\left( -\nu
w/2\right)w^{\nu/2 - 1}, \mbox{ for } w > 0,
$%\end{equation*}
is the gamma distribution with shape and rate parameters both given by
$\nu/2$, and denoted by $\text{Gamma}(\nu/2, \nu/2)$. The random variable $w$ can
be understood as a latent characteristic weight. This provides a
hierarchical specification of \eqref{eq:obsmod} that is  defined by
\begin{equation*}
{\mathcal L}(\bm{y}_i^o \given w_{i}, z_{ik} = 1) \equiv
\Norm_{p_i^o}\left( \bm{O}_i\bm{\mu}_k,
\frac{1}{w_{i}}\bm{O}_i\bm{\Sigma}_k\bm{O}_i'\right), \quad {\mathcal
  L}(w_{i} \given z_{ik} = 1) \equiv \Gam\left(\nu_k/2, \nu_k/2\right), \quad
\bm{z}_i \sim \text{Multinomial}\left(1; \pi_1, \dots, \pi_K\right)
\end{equation*}
where $\bm{z}_i = (z_{i1}, \dots, z_{iK})'$ is the vector of latent class labels and $w_i$ is the characteristic weight for the $i^\text{th}$ observation. 

Together, we take $\{\bm{y}_i^o, \bm{z}_i, w_i: i = 1,\dots,n \}$ to
be the complete data, disregarding the missing values
$\{\bm{y}_i^m: i = 1 ,\dots, n\}$. The corresponding complete data log
likelihood for the parameters $\bm{\Theta}$ is, but for an additive
constant, given by
\begin{align}
\ell_c\left(\bm{\Theta}\given \bm{y}^o, \bm{w}, \bm{z}\right) 
&= \sum_{i=1}^n\sum_{i=k}^K \Bigg[ z_{ik}\log \pi_k 
+ z_{ik}\log \Norm_{p_i^o}\left(\bm{y}_i^o;~ \bm{O}_i\bm{\mu}_k, \frac{1}{w_{i}}\bm{O}_i\bm{\Sigma}_k\bm{O}_i'\right) 
+ z_{ik} \log \Gam\left(w_{i};~ \nu_k/2, \nu_k/2\right) \bigg]  \nonumber\\ 
&= \sum_{k=1}^K\sum_{i=1}^n \Bigg[ z_{ik}\log \pi_k 
 - \frac{z_{ik}}{2}\left(\log\lvert\bm{O}_i\bm{\Sigma}_k\bm{O}_i'\rvert
+ w_{i}(\bm{y}_i - \bm{\mu}_k)'\bm{O}_i'(\bm{O}_i\bm{\Sigma}_k\bm{O}_i')^{-1}\bm{O}_i(\bm{y}_i - \bm{\mu}_k) \right) \nonumber\\ 
&\hspace{2cm} + z_{ik} \left(
\frac{\nu_k}{2}\log\left(\frac{\nu_k}{2}\right) - \log\Gamma\left(\frac{\nu_k}{2}\right) + \frac{\nu_k}{2}\left(\log w_{i} - w_{i}\right) \label{eq:ll}
\right) \bigg].
\end{align}
\subsubsection{An AECM algorithm for parameter estimation} \label{sec:em}
We now design an AECM algorithm for maximum likelihood estimation of
all the model parameters $\bm{\Theta} = \{\bm{\mu}_1, \bm{\Sigma}_1,
\nu_1, \dots, \bm{\mu}_K, \bm{\Sigma}_K, \nu_K\}$ assuming the
complete data log likelihood function in \eqref{eq:ll}. The AECM
approach differs from a general EM approach in that it breaks down each
iteration by partitioning the parameter space into blocks, and cycles
through the partition by alternating between updating each block of parameters
through a conditional maximization (CM-step) and an E-step. For fully
observed data, our 
approach reduces to that of \citet{andrews18} for unconstrained dispersion matrices.

The Q-function, given previous iteration parameter estimates $\hat{\bm{\Theta}}$, is 
$%\begin{equation*}%\label{eq:qfunc}
Q\left(\bm{\Theta} \given \hat{\bm{\Theta}}\right)
= \sum_{k=1}^k \bigg[ 
Q_1\left(\pi_k \given \hat{\bm{\Theta}}\right) + 
Q_2\left(\bm{\mu}_k, \bm{\Sigma}_k \given \hat{\bm{\Theta}}\right) + 
Q_3\left(\nu_k \given \hat{\bm{\Theta}}\right)
\bigg],
$%\end{equation*}
where (excluding constant terms) we have
\begin{align*}
Q_1\left(\pi_k \given \hat{\bm{\Theta}}\right) 
&= \sum_{i=1}^n \hat{z}_{ik} \log \pi_k, \\
Q_2\left(\bm{\mu}_k, \bm{\Sigma}_k \given \hat{\bm{\Theta}}\right)
&= \sum_{i=1}^n \frac{\hat{z}_{ik}}{2}
\bigg[- \log\lvert\bm{O}_i\bm{\Sigma}_k\bm{O}_i'\rvert 
- \hat{w}_{ik}(\bm{y}_i - \bm{\mu}_k)'\bm{O}_i'(\bm{O}_i\bm{\Sigma}_k\bm{O}_i')^{-1}\bm{O}_i(\bm{y}_i - \bm{\mu}_k)  \bigg], \\
\mbox{ and }\qquad Q_3\left(\nu_k \given \hat{\bm{\Theta}}\right)
&= \sum_{i=1}^n \hat{z}_{ik} \bigg[
\frac{\nu_k}{2}\log\left(\frac{\nu_k}{2}\right)
-\log\Gamma\left(\frac{\nu_k}{2}\right) 
 + \frac{\nu_k}{2} \left\{ \log \hat{w}_{ik}
+ \psi\left( \frac{\hat{\nu}_k + p_i^o}{2}\right) - \log\left( \frac{\hat{\nu}_k + p_i^o}{2}\right) - \hat{w}_{ik} \right\}
\bigg],
\end{align*}
with $\psi(\cdot)$ as the digamma function, $\hat{z}_{ik}$ as the
(current iteration) posterior probability that the $i$th record belongs to cluster $k$, and $\hat{w}_{ik}$ is the (current iteration) conditional expectation of $w_i$ given $\bm{y}_i^o$ and $z_{ij}=1$. The later weights the influence of $\bm{y}_i^o$ in estimation of $\bm{\mu}_k$ and $\bm{\Sigma}_k$. 

In the E-step, given current parameter estimates $\hat{\bm{\Theta}}$, we obtain the updates
\begin{align*}
\hat{z}_{ik} & \equiv  \E_{ \hat{\bm{\Theta}} } \left( z_{ik} \given
  \bm{y}_i^o \right)
  = \frac{ \hat{\pi}_k t_{p_i^o}\left(\bm{y}_i^o;~
  \bm{O}_i\hat{\bm{\mu}}_k, \bm{O}_i\hat{\bm{\Sigma}}_k\bm{O}_i',
  \hat{\nu}_k\right) } { \sum_{k'=1}^K \hat{\pi}_{k'}
  t_{p_i^o}\left(\bm{y}_i^o;~  \bm{O}_i\hat{\bm{\mu}}_{k'},
  \bm{O}_i\hat{\bm{\Sigma}}_{k'}\bm{O}_i', \hat{\nu}_{k'}\right) },
\quad\mbox{ and } \\
\hat{w}_{ik} \equiv & \E_{ \hat{\bm{\Theta}} } \left( w_i \given \bm{y}_i^o, z_{ik} = 1 \right)
= \frac{ \hat{\nu}_k + p_i^o }{\hat{\nu}_k + (\bm{y}_i - \hat{\bm{\mu}}_k)'\bm{O}_i'(\bm{O}_i\hat{\bm{\Sigma}}_k\bm{O}_i')^{-1}\bm{O}_i(\bm{y}_i - \hat{\bm{\mu}}_k)}.
%& \E_{ \hat{\bm{\Theta}} } \left( \log w_i \given \bm{y}_i^o, z_{ik} = 1 \right) 
%= \log \hat{w}_{ik} + \psi\left( \frac{\hat{\nu}_k + p_i^o}{2}\right) - \log\left( \frac{\hat{\nu}_k + p_i^o}{2}\right)
\end{align*}

To define the CM-steps, we form the parameter space partition
$\bm{\Theta} = \big\{ \{\pi_1, \bm{\mu}_1, \nu_1 \dots, \pi_K,
\bm{\mu}_K, \nu_K\},~\{\bm{\Sigma}_1, \dots,\bm{\Sigma}_K\}\big\}$
following \citet{andrews12}. Our computation in the CM-steps makes use
of missingness indicator vectors $$\bm{a}_i = \big[ \I(y_{i1} \text{ is
  observed}), \dots, \I(y_{ip} \text{ is observed})\big]'$$ defined for
$i = 1, \dots, n$. In the first CM-step, we update the $\pi_k$'s according to
%	\begin{equation*} %\label{eq:cmpi}
$	\hat{\pi}_k = { \sum_{i=1}^n \hat{z}_{ik}}/{n} $,
%	\end{equation*}
	where the numerator $\sum_{i=1}^n \hat{z}_{ik}$ can be
        understood as representing an estimated (current iteration)
        sample size from  the $k$th group. Also in the first CM-step,
        for computational simplicity, we can
        update the $\bm{\mu}_k$s, following derivations in 
        Appendix~\ref{appendixa}, by  
	\begin{equation}\label{eq:cmmu}
	\hat{\bm{\mu}}_k = \left(\sum_{i=1}^n\hat{z}_{ik}\hat{w}_{ik}\diag(\bm{a}_i)\right)^{-1}\sum_{i=1}^n\hat{z}_{ik}\hat{w}_{ik}\diag(\bm{a}_i)\bm{y}_i,
	\end{equation}
        if it increases $Q_2(\cdot)$, or we keep the current
        value (however, see the comment below). Compared to  updates
        for $\bm{\mu}_k$ using fully         observed data, the missingness indicators
        $\bm{a}_i$ in the RHS of \eqref{eq:cmmu}      only add
        element-wise contributions for observed values.  Similarly, the
        LHS term serves as an element-wise number-of-observations
        adjustment for the number of observed values in each
        feature. The first CM-step then updates the $\nu_k$s as the solution to
 	\begin{equation} \label{eq:cmnu}
  1 + \log\left(\frac{\nu_k}{2}\right) + 1 -
  \psi\left(\frac{\nu_k}{2}\right)  +
  \frac{1}{\sum_{i=1}^n\hat{z}_{ik}}\sum_{i=1}^n \hat{z}_{ik} \bigg[
  \log \hat{w}_{ik}+ \psi\left( \frac{\hat{\nu}_k + p_i^o}{2}\right) -
  \log\left( \frac{\hat{\nu}_k + p_i^o}{2}\right) - \hat{w}_{ik}
  \bigg] = 0.
 	 \end{equation}
There is no closed-form solution to~\eqref{eq:cmnu}. Our {\tt R} package {\sc MixtClust} offers a numerical solution using Brent's method \citep{brent71} and, by default, also extends the closed-form approximation introduced by \citet{andrews18} that uses
  	\begin{equation*}%\label{eq:approxdf}
	\hat{\nu}_k \approx \frac{-\exp(v_k) + 2\exp(v_k)\left[ \exp\left( \psi\left(\frac{\hat{\nu}_k^\text{old}}{2}\right)\right) - \frac{\hat{\nu}_k^\text{old}}{2} + \frac{1}{2}\right]}{1-\exp(v_k)},
	\end{equation*}
  	by modifying $v_k$ to only use the observed number of features $p_i$, in contrast to $p$, for the $i^\text{th}$ observation: 
	\begin{equation*}%\label{eq:approxdfcont}
	v_k = -1 - \frac{1}{\sum_{i=1}^n \hat{z}_{ik}}\sum_{i=1}^n\left[\hat{z}_{ik}(\log\hat{w}_{ik} - \hat{w}_{ik}) - \psi\left(\frac{\hat{\nu}^\text{old}_k + p_i}{2}\right) + \log\left(\frac{\hat{\nu}^\text{old}_k + p_i}{2}\right)\right].
	\end{equation*}
	
	The second CM-step updates (see Appendix~\ref{appendixa} for
        derivations and 
        explanations), also for computational reasons, suggests updating    $\bm{\Sigma}_k$'s  as
  	\begin{equation}\label{eq:cmsig}
 	 \hat{\bm{\Sigma}}_k = 
 	 \left( \sum_{i=1}^n \hat{z}_{ik} \bm{a}_i  \bm{a}_i'  \right)^{\odot -1}
  	\odot \left(\sum_{i=1}^n \hat{z}_{ik}\hat{w}_{ik}
  	\diag(\bm{a}_i)(\bm{y}_i - \hat{\bm{\mu}}_k)(\bm{y}_i - \hat{\bm{\mu}}_k)'\diag(\bm{a}_i)\right),
    \end{equation}
    if $Q_2(\cdot)$ is increased, or we keep the current value for
    $\hat\Sigma_k$ (see, as before, the comment below).
  where $(\cdot)^{\odot - 1}$ denotes Hadamard (element-wise) inverse,
  $\odot$ denotes Hadamard product, and $\otimes$ denotes tensor
  product. The missingness indicators play similar roles as with
  $\bm{\mu}_k$s, but here they operate on the
  elements of a $p\times p$ matrix rather than a $p$-vector. Our
  suggested updates therefore provide for a generalized EM
  algorithm~\citep{dempsteretal77,mclachlanandkrishnan08}.
\paragraph*{Comment} The check for the $Q_2(\cdot)$ at each of the two CM
steps can be quite 
expensive, so our implementation instead checks for an increase in the
loglikelihood at each cycle, terminating the algorithm at the previous
iteration if there is a reduction in the total loglikelihood. The
calculation of the loglikelihood is needed to be made at each cycle to
assess convergence, so does not increase computational cost.
 
 \paragraph{Comparison to full EM}%\label{sec:comparison}
 Our approach uses marginalization, therefore excluding consideration
 of the missing data values  $\bm{y}_1^m, \dots, \bm{y}_n^m$ in the
 formulation of the incomplete dataset for EM style algorithms and so
 does not utilize the distribution of missing values conditional on
 the observed values, i.e., $f\left(\bm{y}_i^m \given \bm{y}_i^o,
   \bm{\Theta}\right)$, which contrasts to the approach of
 \citet{lin2014}. While their approach uses a different parameter space
 partition, the primary difference is that their CM-step updates for
 $\bm{\mu}_k$ and (general covariance structure) $\bm{\Sigma}_k$
 replaces~\eqref{eq:cmmu} and \eqref{eq:cmsig} by
$$
\hat{\bm{\mu}}_k =
\frac{\sum_{i=1}^n \hat{z}_{ik}\hat{w}_{ik} \hat{\bm{y}}_{ik}}{\sum_{i=1}^n \hat{z}_{ik} \hat{w}_{ik}}, %\label{eq:linmu} \\
\qquad\mbox{and}\qquad \hat{\bm{\Sigma}}_k = \frac{\sum_{i=1}^n  \hat{\bm{\Omega}}_{ik}}{\sum_{i=1}^n \hat{z}_{ik}} %\label{eq:linsig},
$$
where
\begin{align}
\hat{\bm{y}}_{ik} 
&\equiv \E_{\hat{\bm{\Theta}}}\left(
\bm{y}_i \given \bm{y}_i^o, w_{i}, z_{ik} = 1
\right) 
= \hat{\bm{\mu}}_k + \hat{\bm{\Sigma}}_k \bm{O}_i'(\bm{O}_i \hat{\bm{\Sigma}}_k\bm{O}_i')^{-1}\bm{O}_i (\bm{y}_i - \hat{\bm{\mu}}_k) \label{eq:yhat} \\
\hat{\bm{\Omega}}_{ik}
&\equiv \E_{\hat{\bm{\Theta}}}\left(
z_{ik}w_{i}(\bm{y}_i - \bm{\mu}_k)(\bm{y}_i - \bm{\mu}_k)' \given \bm{y}_i^o
\right) 
= \hat{z}_{ik}\left[
\hat{w}_{ik}(\hat{\bm{y}}_{ik} - \hat{\bm{\mu}}_k)(\hat{\bm{y}}_{ik} - \hat{\bm{\mu}}_k)'
+ \big(\bm{I}_p - \hat{\bm{\Sigma}}_k \bm{O}_i'(\bm{O}_i \hat{\bm{\Sigma}}_k\bm{O}_i')^{-1}\bm{O}_i\big)\hat{\bm{\Sigma}}_k
\right] \label{eq:omegahat}
\end{align}
are updated in the E-step. %Because our calculations are only on the
% observed parts of the data,
Our approach only performs computations on the
$\sum_{i=1}^n\sum_{j=1}^p a_{ij} \le  np$ observed values in the
dataset, rather than on all $n p$ values, in the updates for the 
$\bm{\mu}_k$'s and $\bm{\Sigma}_k$'s in this full EM approach. We also
avoid computing ~\eqref{eq:yhat} and \eqref{eq:omegahat} altogether,
which in the full EM approach need to be updated between every
CM-step. While imputation  approaches also avoid evaluating the
equations, they perform computations on $n p$ values. Therefore, our
method by design has fewer computations and is faster than all
comparative methods that account for missing values.   However,
our algorithm is also prone to premature termination, with effects
that we account for in the next section. 
\subsubsection{Initialization and convergence assessment} \label{sec:conv}
EM algorithms and their variants such as AECM find solutions in the
vicinity of their initialization, with convergence leading to the
discovery of a local, but not necessarily global,
solution. Consequently, good starting values are important for its
good performance. \citet{biernackietal03} provided the em-EM algorithm
that runs the EM algorithm (``em'' or ``short EM'') from multiple starting points to lax
convergence and then chooses the solution with the highest loglikelihood and
runs the EM algorithm to strict convergence. 
A modification~\citep{maitra13} proposes eschewing the potentially
expensive loglikelihood check at each ``em'' step, running each ``em''
attempt for a small fixed number of iterations, and then runs the
top few ones with highest loglikelihood for the long run and then
choosing the one that upon convergence the estimate produces the
largest log likelihood. We use this ``Modified em-EM''
approach~\citep{maitra13} in our AECM algorithm for the full EM computations, 
with $K\sqrt{np}$ short runs and 10 long runs. However, our
marginalization approach does not guarantee increases in the
loglikelihood, so we adopt the Rnd-EM algorithm of
\citet{maitra09}, that eliminates the ``em'' steps
altogether. We address the case of premature termination by having a
large number (here, 10, to allow for fair comparison across methods,
but perhaps much more) of long runs. We choose the initial seeds in the same manner as
\citet{lithio18} and follow with one run of the EM cycle.
% a large number of short runs of the EM, or in this case AECM,
% algorithm are performed beginning at random locations throughout the
% parameter space. The locations which attained the highest log
% likelihood are subsequently selected for use in a long run of the
% AECM algorithm until convergence. ~\citep[The modification proposed
% by][takes a few locations for the long runs and then chooses the one
% with that upon convergence the estimate produces the largest log
% likelihood.]{maitra13} By default, we use $10npK$ starting locations
% and perform five AECM iterations in the short runs, then select the
% top four to advance to the long runs.
Beyond checking for decreases in the loglikelihood (for the
marginalization case), we use a lack-of-progress criterion to assess algorithm convergence, stopping when $\ell\big(\hat{\bm{\Theta}}^{(t+1)}\big) - \ell\big(\hat{\bm{\Theta}}^{(t)}\big) < \varepsilon$ for a desired small $\varepsilon > 0$ (we use $\varepsilon = 0.001$), where $\ell$ is the observed-data log likelihood resulting from \eqref{eq:obsmod} and $\hat{\bm{\Theta}}^{(t)}$ is the estimate of $\bm{\Theta}$ at the $t^\text{th}$ AECM iteration.

\subsection{Determining the number of clusters} \label{sec:chooseK}
The AECM algorithm presented here assumes a known $K$ or number of clusters. However, this is rarely the case in applications, and we precede by formulating the choice of $K$ as a model selection problem. Commonly, the Bayesian information criterion (BIC) \citep{schwarz78} is used to discriminate between competing models for a given dataset, and is defined by
%\begin{equation} %\label{eq:bic}
$\text{BIC} = -2\ell\big(\hat{\bm{\Theta}}\big) + m \log n,$
%\end{equation}
where $\ell\big(\hat{\bm{\Theta}}\big) $ is the maximized
observed-data log likelihood and $m$ is the number of free parameters
to be estimated. We choose $K$  by choosing, from among a range of
candidate values, the one attaining the smallest BIC at
convergence. This approach of using BIC to determine the
number of clusters in finite mixture models is well-established
(see, {\em e.g.}~\citep{andrewsetal11, fraleyandraftery06, melnykovandmaitra10}).

\section{Performance Assessment} \label{sec:sim}
This section reports performance evaluations in simulation experiments
of our methodology relative to competing methods.
\subsection{Simulation study design}\label{sec:simstudy}
We simulated data with varying clustering complexities determined by
the generalized overlap of and maximum eccentricity of
\citep{maitraandmelnykov10} and implemented in the {\tt R} package
{\sc MixSim} \citep{melnykov2012mixsim}. The generalized
overlap~\citet{melnykovandmaitra11}, denoted here by
$\mathring\omega$, adopted ideas from \citet{maitra10} to arrive at a
one-point  measure of clustering complexity, specifically, a numerical
summary of the overall overlap between pairs of clusters. Higher
values of the generalized overlap correspond to increased cluster
overlap and consequently higher clustering complexity.  Eccentricity
controls the shape of the clusters, and is specified as $e = \sqrt{1 -
  d_{\min}/d_{\max}}$, where $d_{\min}$ and $d_{\max}$ correspond to
the smallest and largest eigenvalues of the dispersion matrix. Taking
values in $[0,1]$, a perfect hypersphere has eccentricity $e = 0$
whereas a perfect hyperplane has eccentricity $e = 1$. Our simulations
considered two clustering complexities: low ($\mathring\omega =
0.001,e=0.5$) and high ($\mathring\omega = 0.01,
e=0.9$). For each complexity
level, we simulated 100 complete datasets with $p = 3$ features,  $K
= 3$ clusters, $n = 100$ cases. %(We also tried other amounts of
                              %missingness, but the general results
                              %are similar so we only report results
                              %for $\lambda=0.1$.)
The degrees of freedom were set to be
$\nu_k = 15$ for all %$k = 1,2,3$
three clusters in our simulation
experiments. 

\begin{figure}[h]
\centering
% n=100_p=3_k=3_omega=0.01_lambda=0.1_nu=15_nsim=1
\includegraphics[width=.75\linewidth]{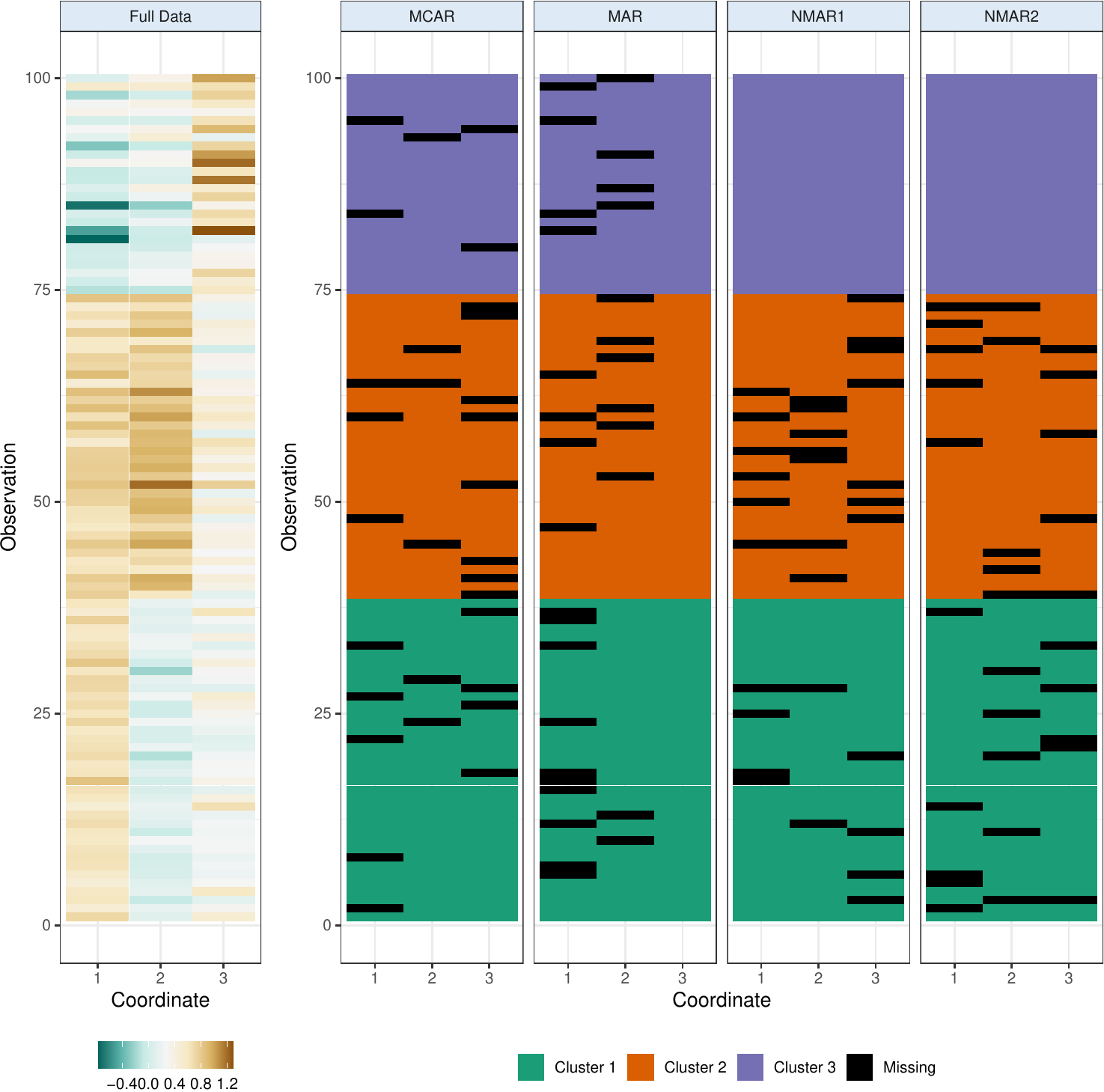}
\caption{Representative sample simulated three-dimensional dataset in
  a high clustering complexity scenario. The leftmost figure is the
  heatmap of a {\sc MixSim}-simulated three-groups full dataset before
  observations were deleted. The right panel of figures provides the
  corresponding   observation status (either missing, or if observed, cluster
  membership) of each observation in the four missingness mechanisms
  (MAR, MCAR, NMAR1, NMAR2).}
\label{fig:simdat}
\end{figure}

Given each full synthetic dataset, we deleted values according to the
MCAR, MAR, NMAR1, and NMAR2 mechanisms
to produce partially recorded datasets for comparing the competing
clustering algorithms. For the MCAR setting, we randomly removed $\lambda np$ values across the
entire dataset so that each element $y_{ij}$ had the same probability
of being missing for $i = 1,\dots, n$ and $j =1, \dots, p$. Under MAR,
we randomly removed $\lambda np$ values in only the first two features
according to MCAR, as in~\citet{chi2016k}. Our experiments considered two versions of NMAR. The NMAR1
version of~\citet{lithio18} had
one cluster being fully observed with the other two
clusters having MCAR observations. The NMAR2 version~\citep{chi2016k,lithio18} also had one cluster fully
observed, but the other clusters have the appropriate bottom quantile
of each feature removed so as to achieve an overall missingness level
$\lambda$. Figure~\ref{fig:simdat} represents a sample dataset in the
high clustering complexity scenario with the four patterns of missingness,
for $\lambda = 0.1$.
(For our experimental evaluations, we ensured that there
were at least $(p+1)$ complete records from each cluster, to
allow for methods such as  deletion to work.) For the first set of experiments, where our comparisons
also included imputation methods, we set $\lambda=0.1$. For the more thorough, comprehensive
evaluations that only considered model-based clustering methods, we
used $\lambda\in\{0.05, 0.1, 0.15, 0.2\}$, as well as samples of 
$n=50$ and $n=100$ observations. The smaller sample size was
chosen to provide us with a sense of performance when the average
number of observations per cluster was not particularly high relative
to dimension. This translated to between
10-16\%, 20-32\%, 26-46\% and 34-57\% of observations in our simulated
datasets having at least one incomplete record.

%\clearpage

\subsection{Comparison methods and evaluation}\label{sec:compete}
From now on, for ease of reference, we label  our
proposed approach as ``Observed EM'',  the full EM approach as ``Full
EM'' and the case deletion approach, 
that only makes use of complete cases, as ``Complete Case.''
%All three approaches are implemented in our {\tt R} package {\sc
%MixtClust} and we use the default settings introduced in
%Section~\ref{sec:methods}. Additionally,
We also consider three imputation schemes to produce completed datasets before applying our software: Amelia II, mi, and mice. The Amelia
II approach of \citet{honaker11} assumes the data are multivariate
Gaussian and combines the EM algorithm with bootstrapping to draw from
the posterior of the complete data parameters that are then used for
imputation. In contrast, the mice and mi methods of \citet{buuren11}
and \citet{su11}, respectively, make use of chained equations to
impute missing values. For all imputation approaches, we used default
settings to generate $M = 5$ completed datasets and performed
clustering using the average of the $M$ imputed values.
%We evaluated our approach for both compute speed as well as
%clustering performance. ur first set of evaluations was on the comparative timings of our ``Observed EM'' approach relative to the  ``Full EM'' approach.  We
We evaluated clustering performance by comparing the
the true cluster partition to  that obtained by each method 
at the BIC-determined best $K$ (from now on denoted by $\hat
K$) via the Adjusted Rand index (ARI) \citep{hubert1985comparing}. The
(unadjusted) Rand index \citep{rand71} is a measure of class agreement
taking value in $[0,1]$, with a value of one indicating perfect
agreement. Under random classification, the Rand index has an expected
value greater than zero, reflecting the fact that, by chance alone,
random classification could correctly classify some observations. The
ARI~\citep{hubert1985comparing} is a modification of the Rand index that, in
contrast, has expected value of zero under random classification while retaining the property that a value of one corresponds to perfect classification.
\subsection{Simulation results}\label{sec:simresults}
\subsubsection{Clustering performance}\label{sec:simres}
\begin{figure}
  \centering
  \mbox{
    \subfloat[Recovery of true $K$]{\includegraphics[width=.83\linewidth]{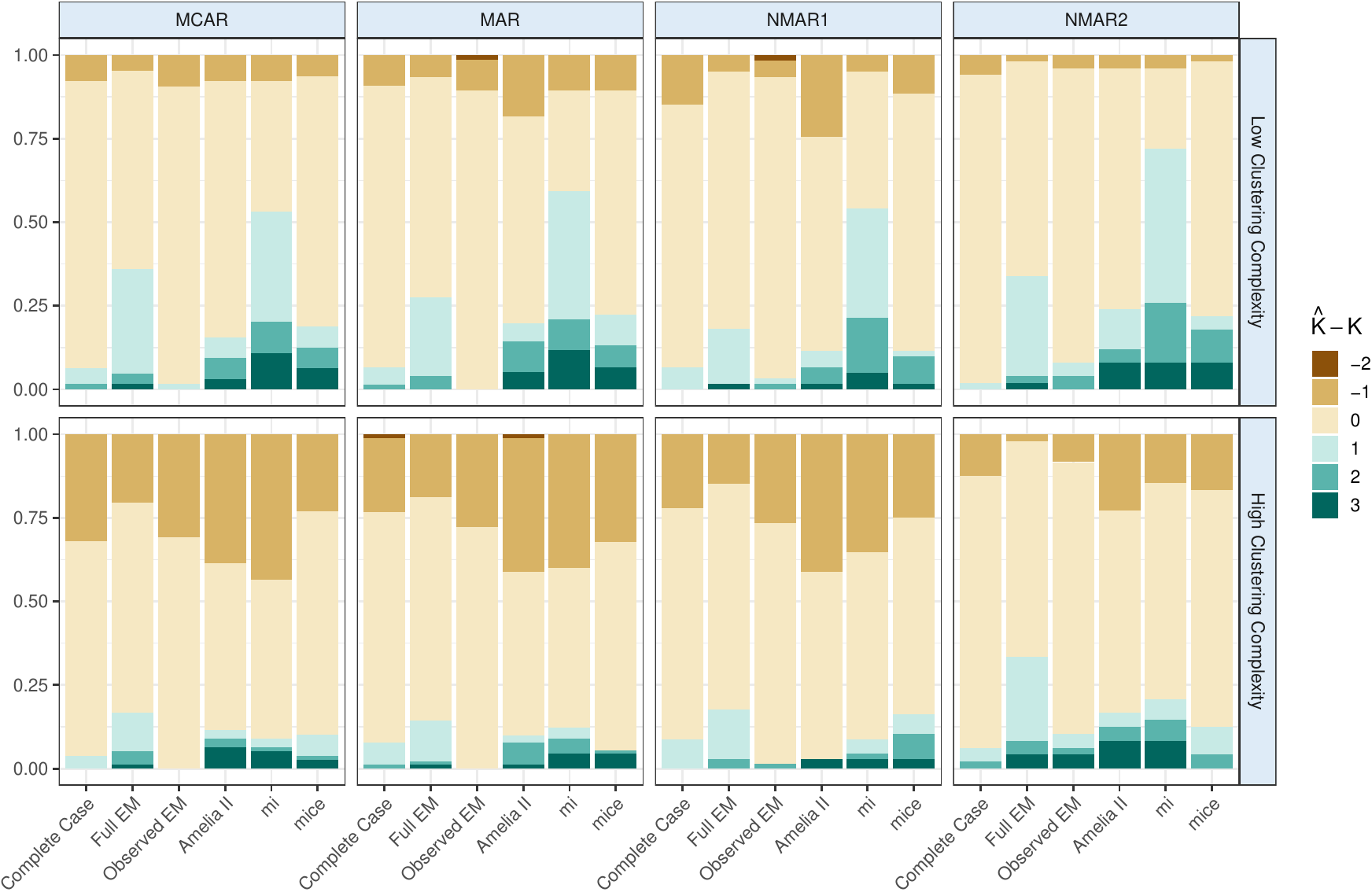}\label{fig:Khat}}}
  \mbox{
  \subfloat[Clustering performance of competing methods, as per ARI]{\includegraphics[width=.83\linewidth]{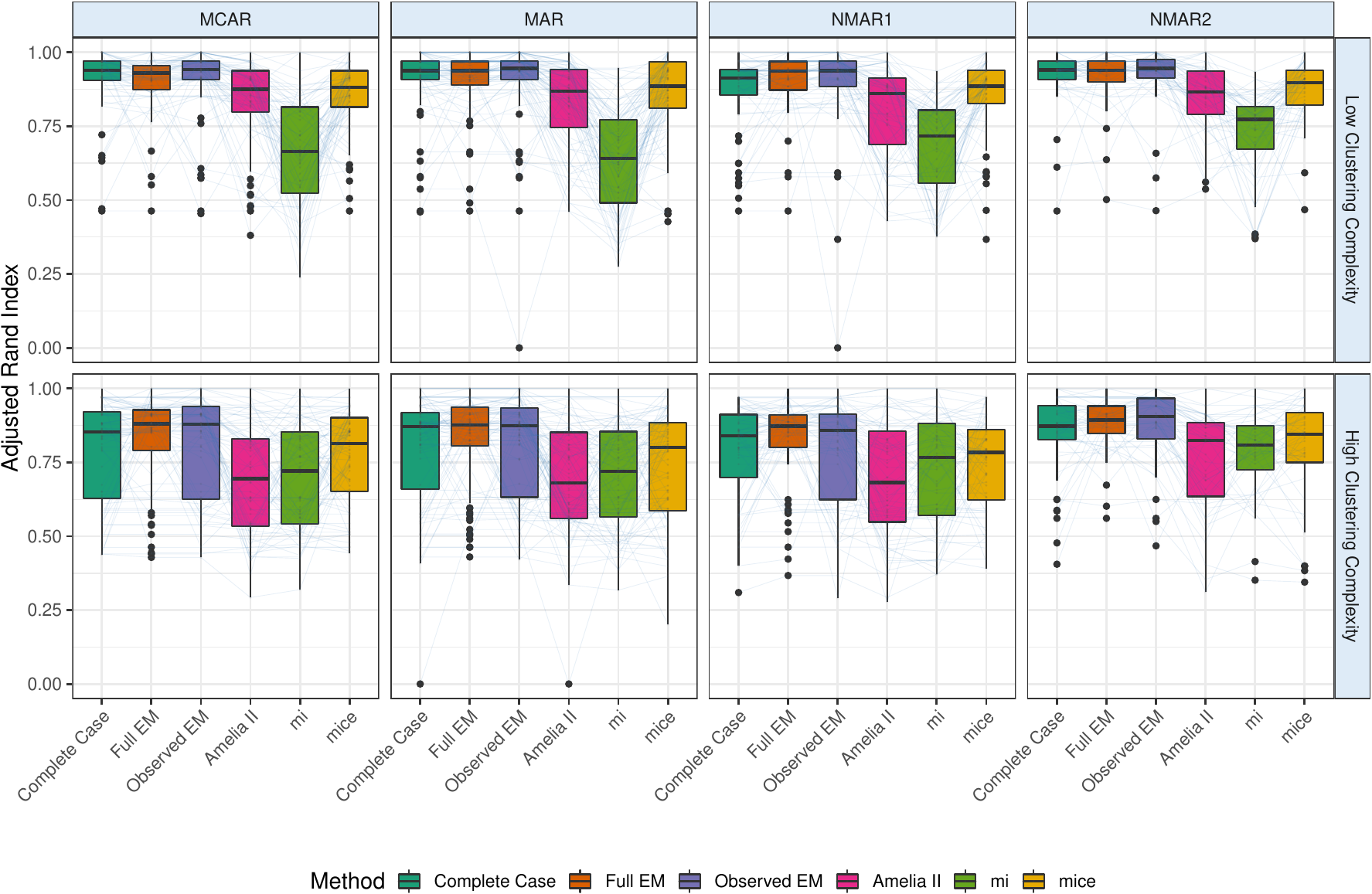}\label{fig:ARI}}}
\caption{Simulation results  across 100 simulation replications for
  each method, missingness mechanism, and clustering complexity
  scenario. (a) accuracy of BIC in recovering the true
  number of clusters demonstrated in terms of the difference between the number
  of clusters   chosen by BIC $\hat{K}$, and the true $K$. (b)
  ARI values comparing the true partition to the clustering
  obtained at $\hat K$.}
% The difference between the number
                                %of clusters   chosen by BIC,
                                %$\hat{K}$, and the true number of
                                %clusters, $K$, is   depicted as a
                                %composition across 100 simulation
                                %replications for    each method,
                                %missingness mechanism, and clustering
                                %complexity   scenario.
%  (b) Adjusted Rand values comparing the true partition to the clustering at the BIC preferred value of $K$ shown as box plots based on 100 simulation replications for each method, missingness mechanism, and clustering complexity scenario.}
\label{fig:sim}
\end{figure}
We evaluate our clustering methods in terms of $K$-selection and in
terms of the ability of our methods to obtain the true
partition. Figure~\ref{fig:sim} displays a summary of our results. We
consider accuracy of BIC in selecting the number of clusters in
Figure~\ref{fig:Khat}.  Under high clustering complexity, most
algorithms tended to select too few, rather than too many,
clusters. However, in general, and except for being edged out by the
``Complete Case'' in NMAR2 for both clustering complexity situations
and MAR for the low clustering complexity scenario, our ``Observed
EM'' algorithm usually deviated 
the least from the true $K$. 
All three imputation methods generally selected more than the true $K$, and this was exaggerated under low clustering
complexity. Within each pattern of missingness and clustering
complexity, ``Observed EM'' provided
competitive recovery of the true $K$. The cluster
partition at $\hat{K}$ was next compared to the true class memberships using
the ARI, and is summarized in Figure~\ref{fig:ARI}.
Overall, our approach produced cluster partitions at least as, or
more, closely aligned with the truth under low clustering complexity
for all missingness mechanisms. On the other hand, for high clustering
complexity problems, our method is only the best overall under NMAR2,
and for the remaining three missingness mechanisms for which the MAR
assumption holds or is not as severely violated, our approach is only
surpassed by ``Full EM''. 

The results of our simulation experiments show good performance of our
``Observed EM'' procedure. In cases with high clustering complexity,
``Full EM'' performs better than our case. However, even here,
``Observed EM'' is quite competitive. We now evaluate both methods
against each other and ``Complete Case'' in a more comprehensive set of
experiments. 

\subsubsection{Comprehensive comparisons between ``Observed EM'',
  ``Full EM'' and ``Complete Case'' methods}
\label{expts:comp}
\paragraph{Evaluations at the true $K$} Our first set of evaluations
assumed that the true $K$ was known. 
\begin{figure}[!h]
  \centering
  \mbox{
    \subfloat[Time to Termination of the three EM algorithms]{\includegraphics[width=\linewidth]{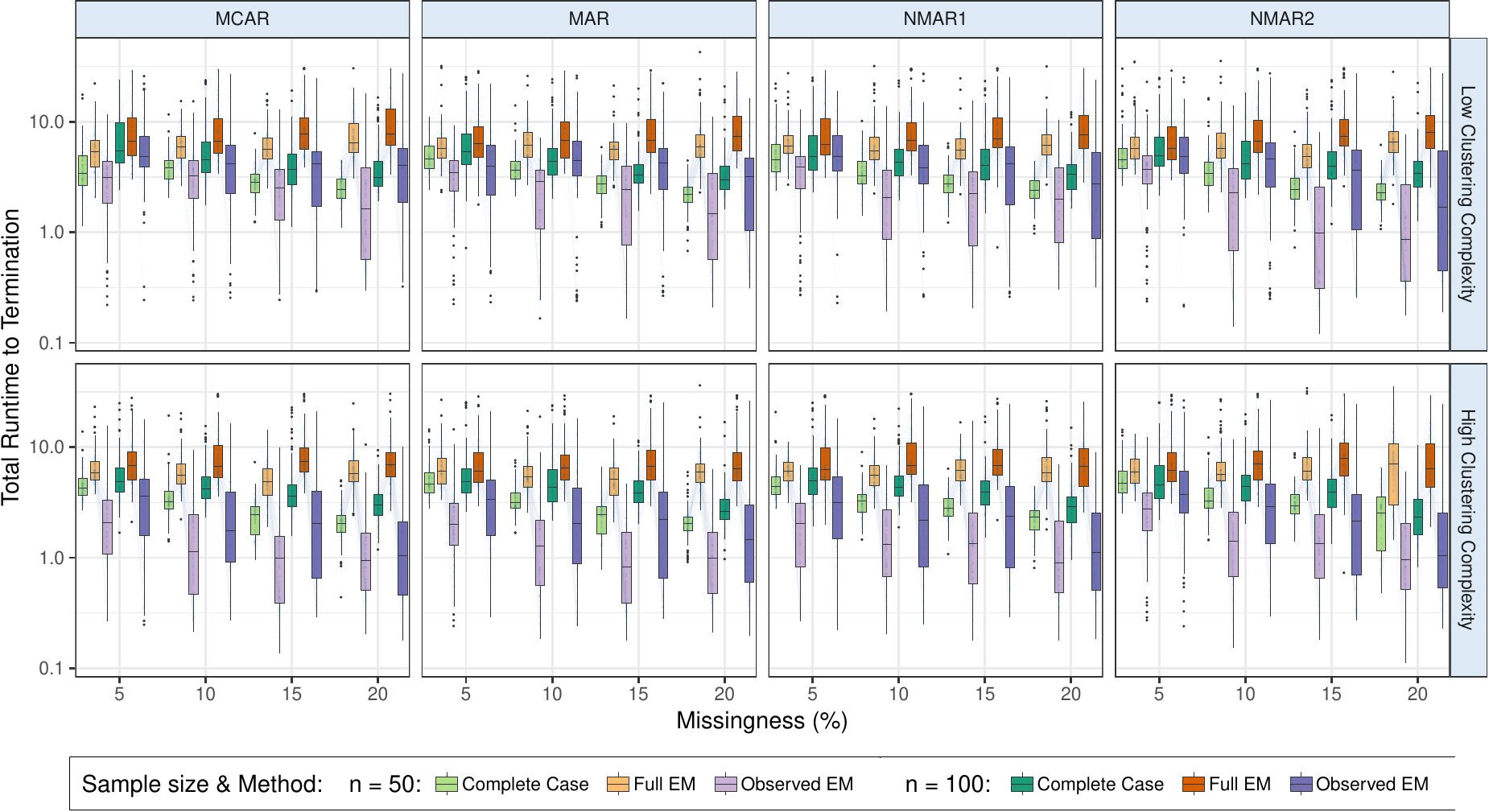}\label{fig:time-fixedk}}}
  \mbox{
  \subfloat[Clustering performance of competing methods, as per ARI]{\includegraphics[width=\linewidth]{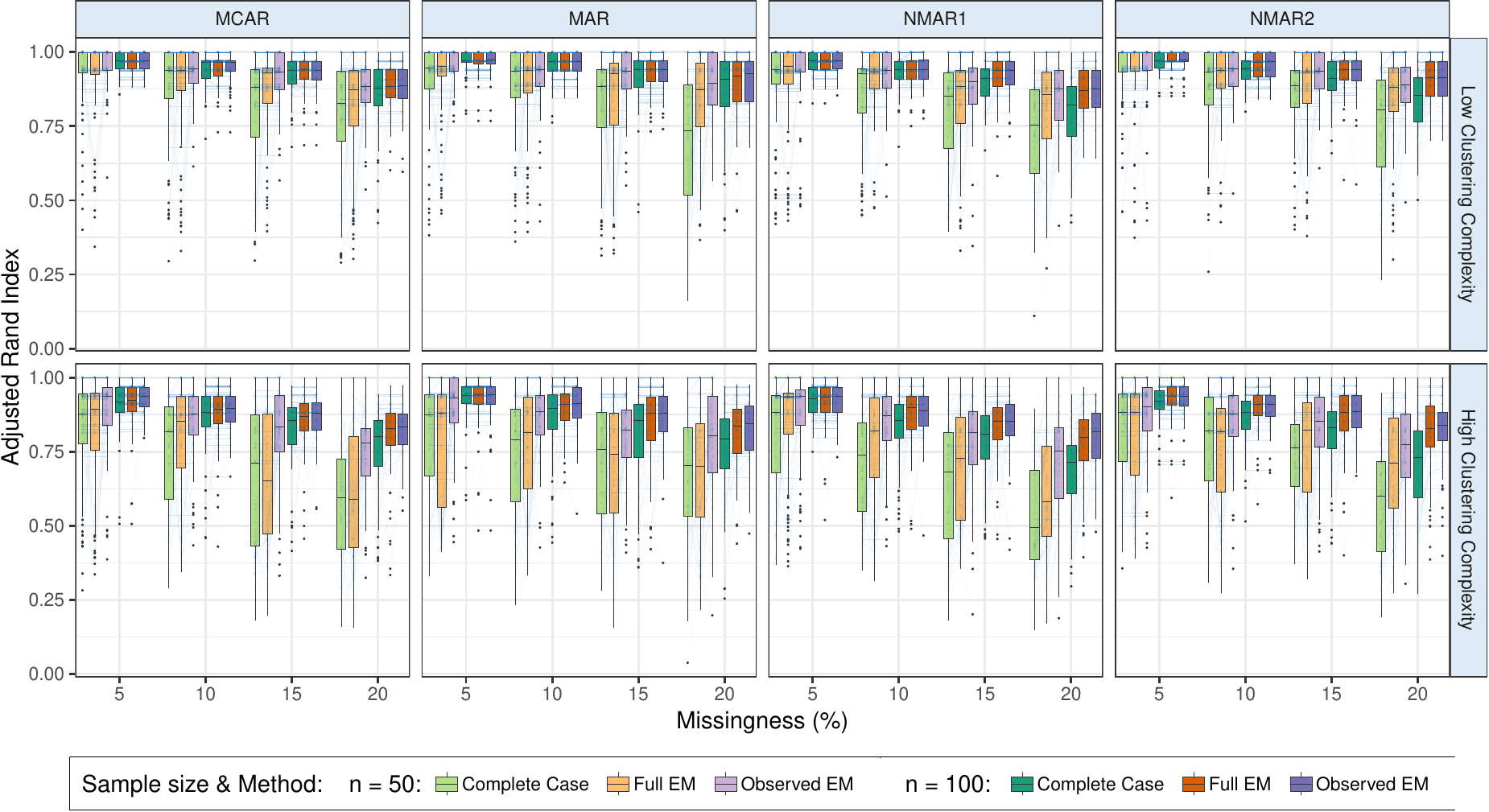}\label{fig:ARI-fixedk}}}
\caption{Comprehensive simulation results  at the known true $K$}
%(a) accuracy of BIC in recovering the true  number of clusters
%demonstrated in terms of the difference between the number   of
%clusters   chosen by BIC $\hat{K}$, and the true $K$. (b)   ARI
%values comparing the true partition to the clustering   obtained at $\hat K$.}
% The difference between the number
                                %of clusters   chosen by BIC,
                                %$\hat{K}$, and the true number of
                                %clusters, $K$, is   depicted as a
                                %composition across 100 simulation
                                %replications for    each method,
                                %missingness mechanism, and clustering
                                %complexity   scenario.
%  (b) Adjusted Rand values comparing the true partition to the clustering at the BIC preferred value of $K$ shown as box plots based on 100 simulation replications for each method, missingness mechanism, and clustering complexity scenario.}
\label{fig:sim-fixedK}
\end{figure}
We report performance for the three mehods in Figure~\ref{fig:sim-fixedK}. 
The discussion at the end of Section~\ref{sec:em} demonstrated that
the number of calculations needed to conduct one iteration of our
``Observed EM'' approach is no more (and often much less) than those
needed for ``Full EM''. However, because the trajectories 
taken by the two methods can be different, we also evaluated the
runtime of each algorithm. Figure~\ref{fig:time-fixedk} shows that the
time taken to termination is almost always smaller for the ``Observed
EM'' case as opposed to the ``Full EM'' case, and this advantage
increases with higher clustering complexity and amounts of
missingness. (Interestingly, the ``Complete Case'' is slower than ``Observed EM'' despite its computations being  often done on
substantially smaller datasets: we attribute this finding to the
difficulty of estimation in smaller datasets.) Our speed also does not
come at a cost to the ability of ``Observed EM'' to recover the true
partition, for Figure~\ref{fig:ARI-fixedk} shows that our method is
quite competitive, and sometimes even slightly better than with ``Full
EM''. An interesting question that arises is whether the choice of
initializer (Modified em-EM for ``Full EM'' but Rnd-EM, by necessity,
for ``Observed EM'') plays a role. Appendix~\ref{appendixb} shows no
appreciable difference between the two initialization approaches on 
the speed and performance of ``Full EM''.

\paragraph{Evaluations with unknown $K$}
Our next set of evaluations required  $K$
to be estimated from $K\in\{1,\ldots,K_{\max}\}$ with $K_{\max}=5,6$ for
$n=50,100$ and for which we used BIC as per Section \ref{sec:chooseK}. 
\begin{figure}[!h]
\centering
% n=100_p=3_k=3_omega=0.01_lambda=0.1_nu=15_nsim=1
\includegraphics[width=\linewidth]{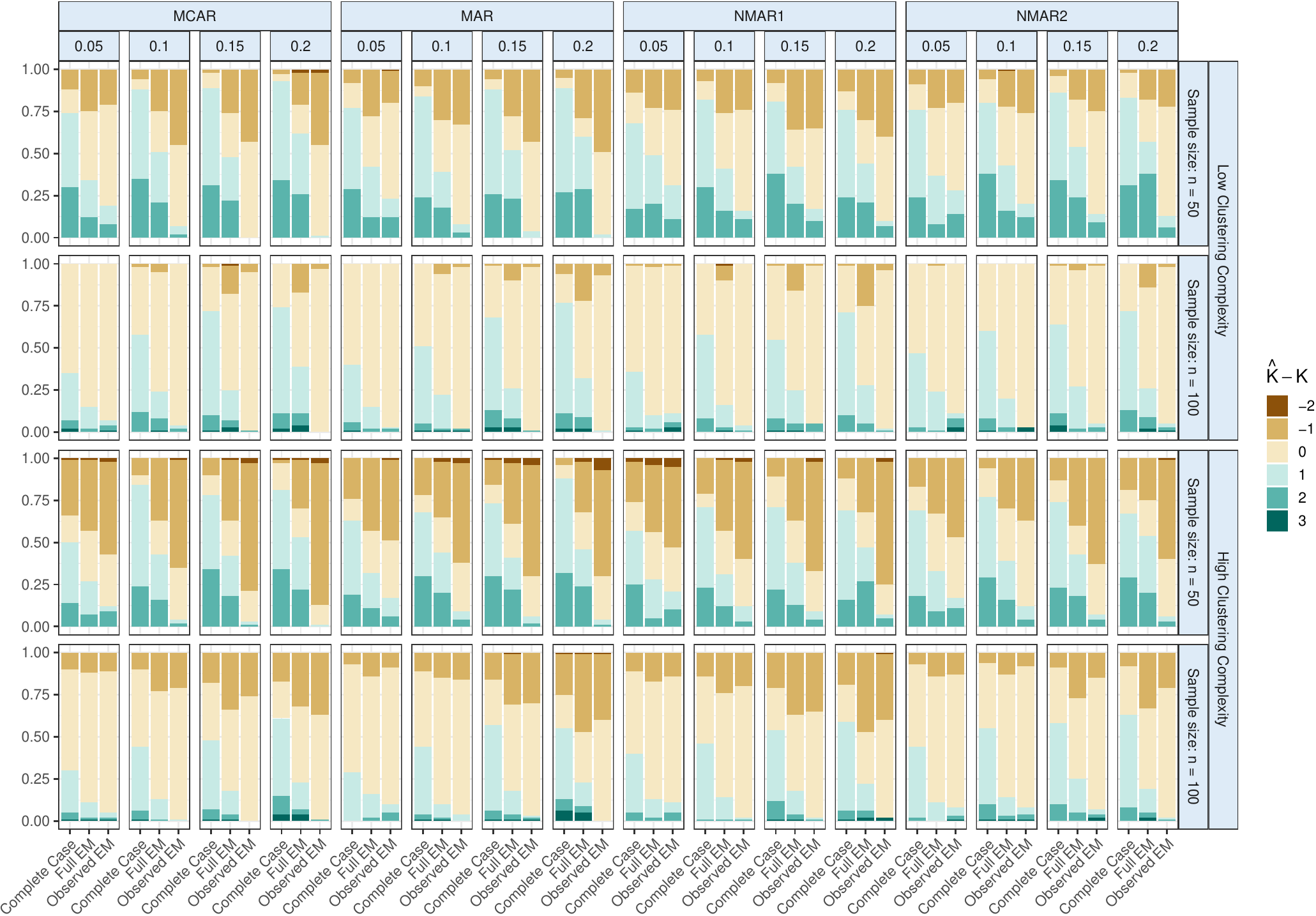}
\caption{Accuracy of BIC, when used with each of the three methods, in
  recovering the true $K=3$, for different clustering complexities,
  sample sizes, missingness mechanisms and $\lambda\in\{0.05,0.1,0.15,0.2\}$.
The display is in terms of the difference between $\hat K$ and $K$. 
}
\label{fig:newKhat}
\end{figure}
Figure~\ref{fig:newKhat} shows relatively good performance of ``Observed
EM'' in estimating $K$, mildly underestimating $K$ for 
  \begin{figure}[!h]
  \centering
  \mbox{
    \subfloat[Time to termination of the EM algorithms in each setting
    when
    estimating $K$ using BIC.]{\includegraphics[width=\linewidth]{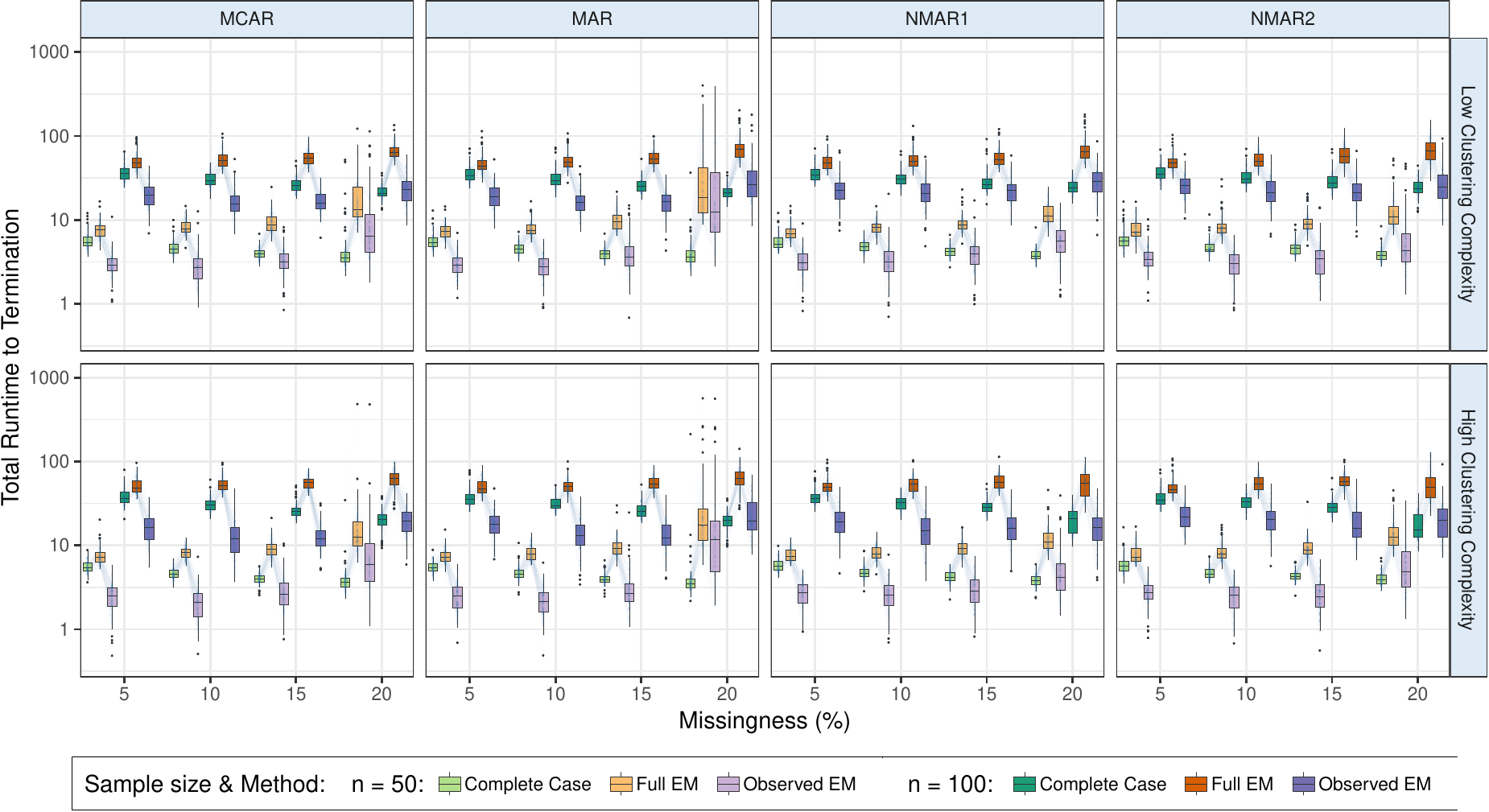}\label{fig:time-estk}}}
  \mbox{
  \subfloat[Clustering performance of competing methods, as per ARI.]{\includegraphics[width=\linewidth]{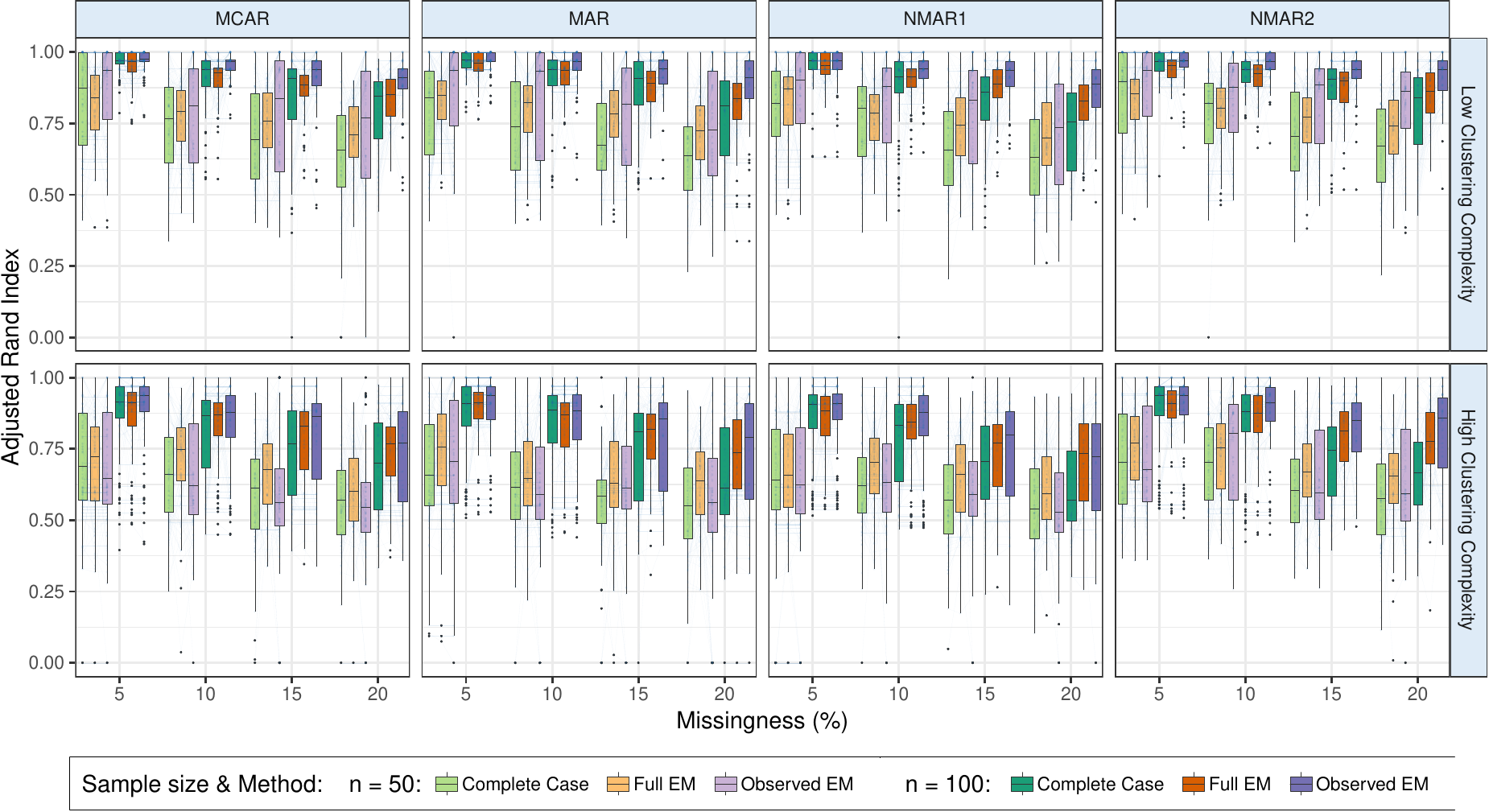}\label{fig:ARI-estk}}}
\caption{Comprehensive simulation results  when $K$ is required to be
  estimated.}
\label{fig:all-estK}
\end{figure}
$n=50$, but not so much for the larger sample size. Performance for
all methods worsens with higher clustering complexity, however
``Complete Case'' and ``Full EM'' moderately overestimate $K$ in all settings. 
As in the known true $K$ case, Figure~\ref{fig:time-estk} shows that ``Observed
EM'' is substantially faster than ``Full EM'' and that, at least in
the cases of low clustering complexity and with larger $n$, is the best overall
performer~(Figure~\ref{fig:ARI-estk}). With high clustering
complexity, this trend is explicit with $n=100$ only for the NMAR2
case, and with lower proportions of missingness 
for  the other three missingness mechanisms. For $n=50$ or with
$n=100$ and MCAR,
MAR, and NMAR1 missingness mechanisms under conditions of high
clustering complexity, the  performance of ``Observed EM'' relative
to ``Full EM'' is more mixed with higher proportions of missingness.

\begin{figure}[!h]
\centering
% n=100_p=3_k=3_omega=0.01_lambda=0.1_nu=15_nsim=1
\includegraphics[width=\linewidth]{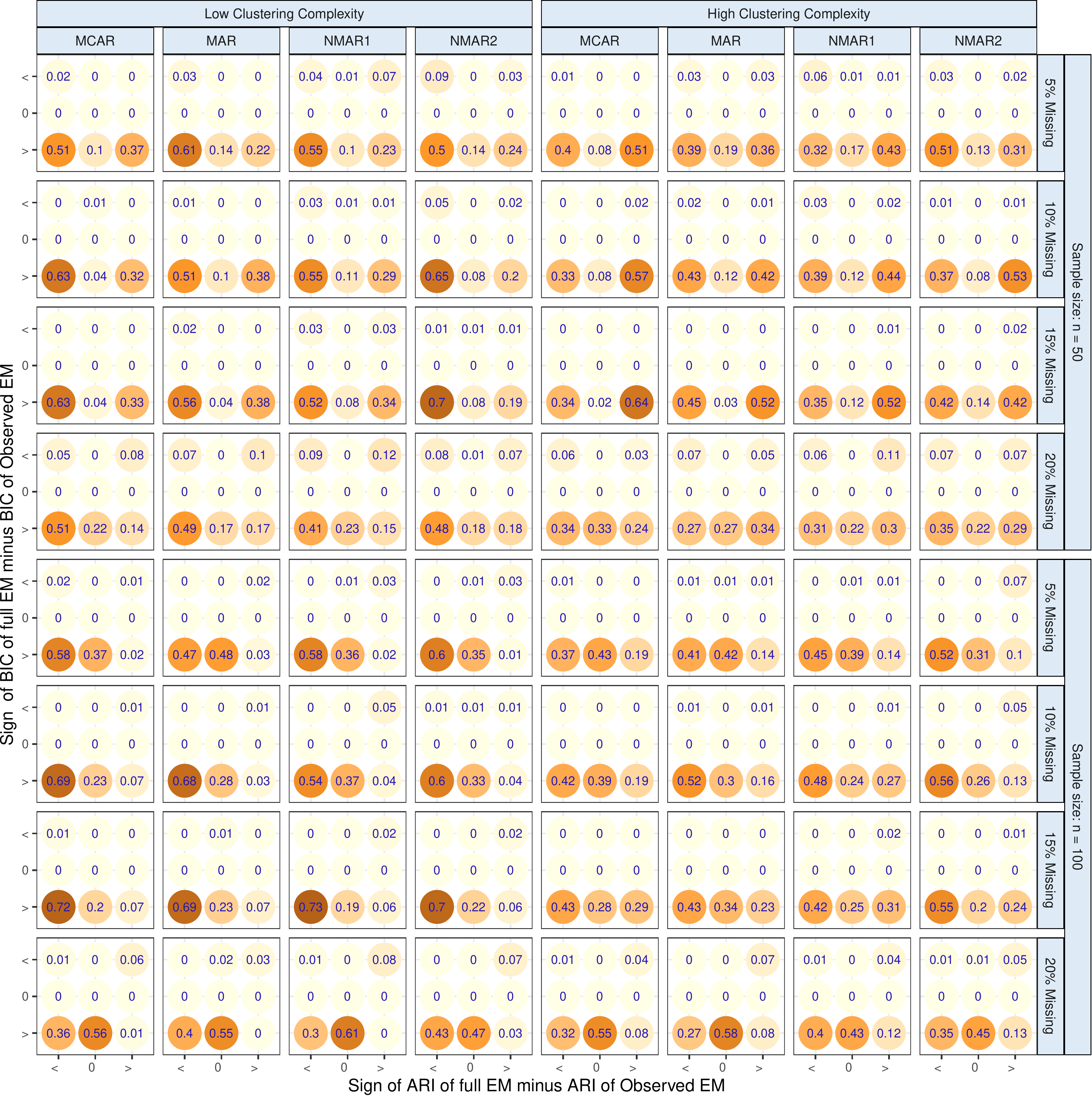}
\caption{BIC relative to ARI for the two methods, for the different
  simulation settings. For each setting, we have cross-tabulated the
  relative frequency of datasets on which where the BIC of ``Full EM''
  is less than (``$<$''), the same (``$0$'') or exceeds (``$>$'') that of
  ``Observed EM'' against the same 
  for the ARI. Color intensity is
proportional to the relative frequency.}
\label{fig:ggtable}
\end{figure}
\paragraph*{Comment}
A reviewer has asked us for insight into the reasons for the better
performance of ``Observed EM'' over ``Full EM'' even when assumptions
are far from MCAR. We surmise the reasons for this. 
Indeed, ``Full EM'' provides guaranteed local
maxima of the EM (AECM) algorithm. Our modification does not,
suffering, as it does, from possible premature termination. However,
the objective log likelihood function, especially in the case of non-MCAR
mechanisms, is not a completely accurate representation of the
sampling mechanism that generated the data, and therefore, maximizing
it may not be the golden bullet that solves the clustering
problem. To evaluate if there indeed is evidence in support of this
hypothesis, we compared the optimal BIC for ``Full EM'' and
``Observed EM'' and the ARI. (The BIC is a surrogate for the goodness
of fit under the assumed loglikelihood model and the ARI is an
indicator of cluster recovery and performance.)
Figure~\ref{fig:ggtable} displays the results. We see that often
``Full EM'' is able to achieve a higher BIC than ``Observed EM'' but
fails to get more credit for cluster recovery (per ARI) and this
mismatch increases correspondingly with proportions of missingness in
data. This phenomenon holds for both when the sample size is
$n=50$ or $n=100$. 
Of course, this does not answer the question as to why
``Observed EM'' does better in terms of cluster recovery. Our hypothesis is that this happens
because our CM-step updates for $\bm{\mu}_k$s and $\bm{\Sigma}_k$s use
method-of-moments estimators and do not condition on the observed
data. When the data are NMAR, the observed values are not directly
informative of those that are missing because their values are related
to their own missingness. We suspect that this aspect in our algorithm
is what is allowing for better performance. Finally, finite
mixture models provide only an indirect approach to clustering with
parameter estimation and model selection happening first, followed by
cluster assignment as a byproduct from the calculation of posterior
probabilities of classification. 
Nevertheless, our performance evaluations establish ``Observed EM''
as a faster and competitive alternative to ``Full EM''. We now apply
it to the problem of characterizing gamma ray bursts.

\section{Discovering the distinct kinds of Gamma Ray Bursts}\label{sec:app}

Gamma-ray bursts (GRBs) are the brightest  electromagnetic bursts
known to occur in space and emanate from distant galaxies. Since
their discovery, several causes of GRBs have been proposed
\citep{chattopadhyay2007statistical,piran05,ackermann2014fermi,gendre2013ultra}
and the existence of multiple sub-types
\citep{shahmoradiandnemiroff15,mazetsetal81,norrisetal84,dezalayetal92}
hypothesized. To
elucidate the origins of GRBs, it is of interest to determine the
number and defining characteristics of these groups. Early work
classified GRBs using one or two features, often using only burst
duration \citep{kouveliotou1993identification}. It was argued that
more variables were needed to fully account for the observed data
structure \citep{feigelsonandbabu98,mukherjeeetal98}, leading to
recent interest in clustering GRBs using more
features. Subsequent
analyses~\citep{chattopadhyay17,chattopadhyayandmaitra18,berryandmaitra19,almodovarandmaitra20} established five groups in the GRB dataset
obtained from the most recent Burst and Transient Source
Experiment (BATSE) 4Br catalog.

 The BATSE 4Br catalog is the most comprehensive database of the duration,
 intensity, and composition of 1,973 GRBs, but the records are subject
 to missing values encoded as zeros
 \citep{chattopadhyay17,chattopadhyayandmaitra18}, leading to a total
 of 1,599 GRBs that are complete cases. There are up to nine features
 for each GRB, namely $T_{50}$, $T_{90}$, $F_1$, $F_2$, $F_3$, $F_4$,
 $P_{64}$, $P_{256}$, and $P_{1024}$, where $T_{\tau}$ denotes the time
 by which $\tau$\% of the flux arrive, $P_{t}$ denotes the peak fluxes
 measured in bins of $t$ milliseconds, and $F_s$ represents the
 fluence in the $s$th spectral channel. Due to the extreme
 right-skewness in the distribution of  these variables, we apply the
 customary base-10 logarithm  transformation to all the variables, and for
 brevity, omit the logarithm in subsequent
 descriptions. (\citep{berryandmaitra19} however incorporated data-driven
 transformations in their analysis to address the skew.)
 The two duration variables $T_{50}$
 and $T_{90}$ are observed for all 1973 GRBs. The three peak flux
 measurements are only missing in one GRB, while $F_1$, $F_2$, $F_3$
 and $F_4$ are missing values in 29, 12, 6 and 339
 GRBs~\citep{chattopadhyayandmaitra18}. Multivariate analysis of the
 GRBs has so far largely focused on the 1599 GRBs with complete
 records~\citep{chattopadhyay17,chattopadhyayandmaitra18,berryandmaitra19,almodovarandmaitra20,feigelsonandbabu98,mukherjeeetal98}.
 On the other hand,~\citet{tothetal19} ignored the peak fluxes and the
 $F_4$ features and  the 44 GRBs that were missing values for the
 other features and performed Gaussian mixture-model-based clustering
 for the 1929 GRBs and came up with three types of GRBs. They also
 tried to explain the results of~\citet{chattopadhyay17} in the context
 of their findings. However, the analysis of
 ~\citet{chattopadhyayandmaitra18} on the complete dataset showed
 all nine variables to have clustering information.
\citet{chattopadhyayandmaitra18} also used the maximum marginal posterior
probability to classify the remaining $n - n'$ incomplete cases to the
obtained cluster partition: however, this approach assumes that all
the clusters are represented in the complete sample, an assumption
that may be untenable, especially when the data are not MCAR. Therefore, it is  of interest to include GRBs with partial records in
our analysis and our development in this paper helps facilitate that investigation. 
 % Here, we apply our proposed approach to these data using nine features: $T_{50}$, $T_{90}$, $F_1$, $F_2$, $F_3$, $F_4$, $P_{64}$, $P_{256}$, and $P_{1024}$, where $T_{\pi}$ denotes the time by which $\pi$\% of the flux arrive, $P_{t}$ denotes the peak fluxes measured in bins of $t$ milliseconds, and $F_s$ represents the fluence in spectral channel $s$. Due to the extreme right-skew of these variables, we apply a base 10 logarithm transformation, and for brevity, omit the logarithm in subsequent descriptions of the variables. 

Since the imputation approaches performed poorly in our simulation
assessments, we restricted our attention to the observed, full, and
complete case EM style approaches, again with default settings. The
``Observed EM'' and ``Full EM'' methods preferred $\hat{K} = 7$
clusters, as per BIC  among 
candidates ranging from $K = 1, \dots, 10$, in contrast to previous
reports of only two or three clusters obtained using a few features or
the five clusters obtained using the complete dataset. Our
methodology, applied on the 1599 complete GRB cases, also found 
five or six groups, though the latter was not conclusive,
following~\citet{kassandraftery95}, so we decide on five groups.
~\citet{chattopadhyayandmaitra18} also got five groups using the 
{\sc teigen} software~\citep{andrews18}. This is reassuring even
though our method differs from that of {\sc
  teigen} insofar as it allows for $\nu$, to vary across groups. We now discuss the results for
the 7-groups solutions (and briefly, for the complete case 5-groups
scenario). 

\begin{figure}[!h]
\mbox{
  \subfloat[]{
    \includegraphics[width=\linewidth]{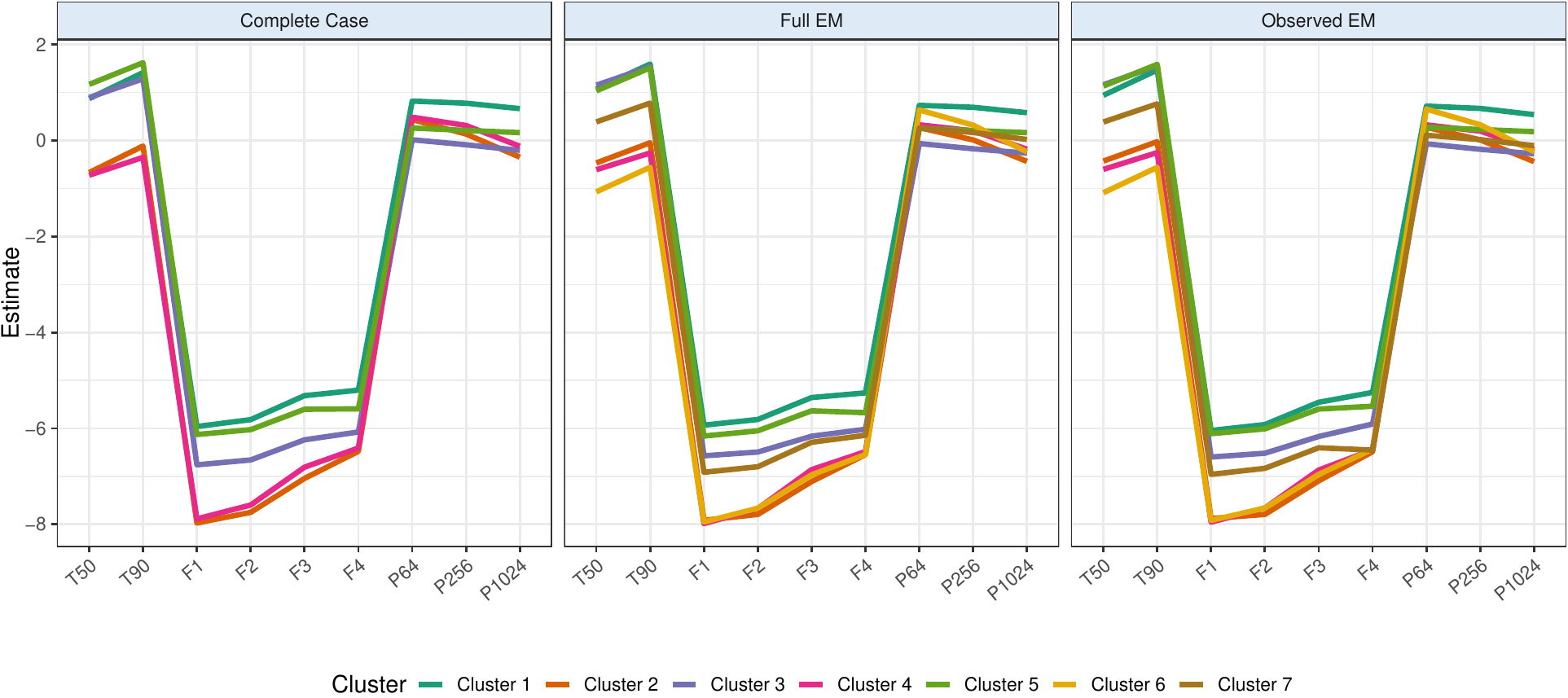}\label{fig:grbmu}
  }}
  \subfloat[]{
    \includegraphics[width=\linewidth]{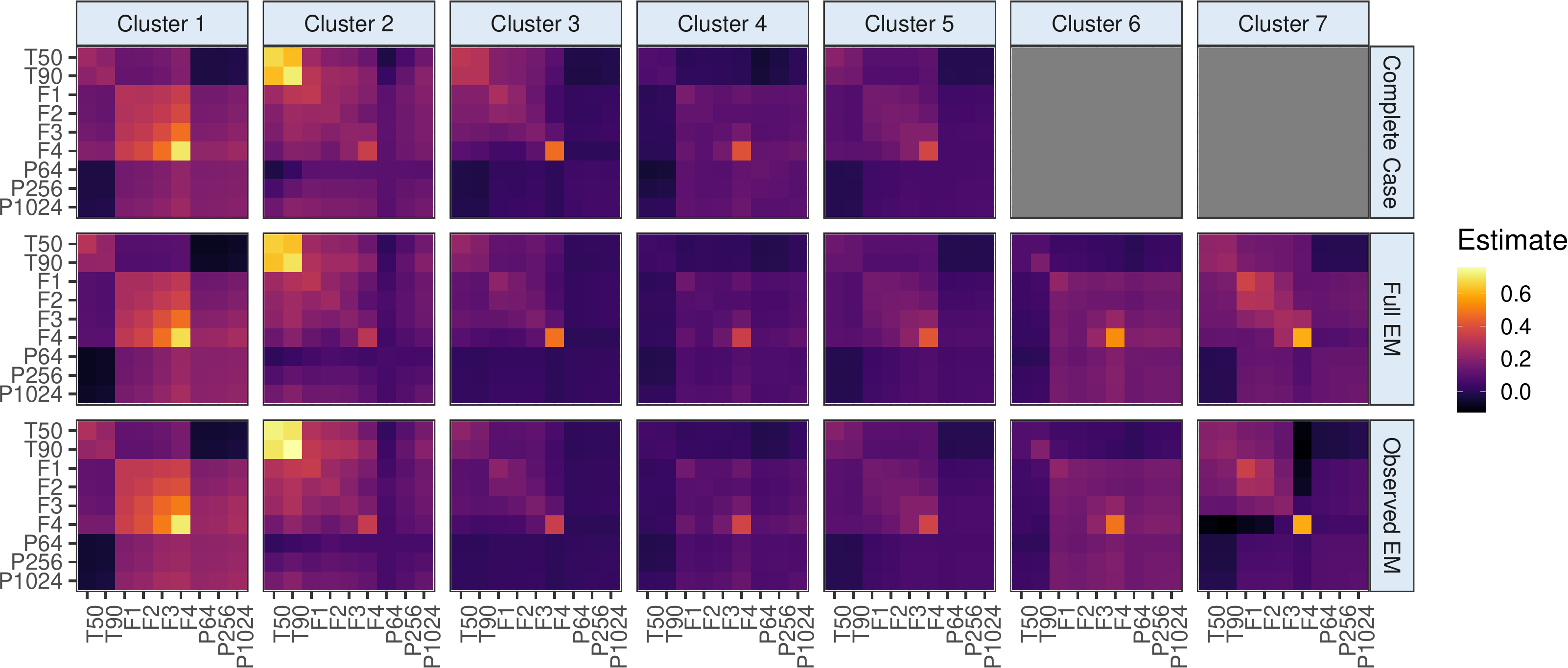}\label{fig:grbsig}
  }
\caption{(a) Parallel coordinate plots of
  $\hat{\bm{\mu}}_1,\ldots,\hat{\bm{\mu}}_{\hat K} $ and (b) heatmap
  of   $\hat{\bm{\Sigma}}_1,\ldots,\hat{\bm{\Sigma}}_{\hat K} $ for the
  BIC-selected $\hat K$ groups partition of the GRB data obtained using each of
  the three methods.}
\label{fig:grb}
\end{figure}

Table~\ref{tab:grbests} presents the estimated cluster proportions and
degrees of freedom for the BIC-preferred K obtained using the three
methods. 
\begin{table}[h!]
  \caption{Estimated (a) mixing proportions $\hat{\pi}_1, \ldots,
    \hat\pi_{\hat {K}}$  and (b) degrees of freedom
    $\hat{\nu}_1,\ldots, \hat{\nu}_{\hat K}$, obtained after fitting
    the three approaches to the GRB dataset at the BIC-preferred $\hat
    K =  5$ (for the complete case) and $\hat K=7$ for the other two cases.}
  \centering
\subfloat[Estimated mixing proportions]{
\begin{tabular}{|r|rrrrrrr|}
  % \multicolumn{7}{c}{Estimated mixing proportions}\\
 \hline
Cluster                     & 1                     & 2
  & 3                     & 4                     & 5
  & 6   & 7              \\ \hline
  Complete Case               & 0.210 & 0.106 & 0.276 & 0.151 & 0.258 & - &- \\
  Full EM                      & 0.191 & 0.086 & 0.246 & 0.083 & 0.187 & 0.081 & 0.126 \\ 
  Observed EM                     &0.204 & 0.079 & 0.215 & 0.096 & 0.214 & 0.079 & 0.113 \\ 
  \hline
  \multicolumn{7}{c}{~}
\end{tabular}
}
\subfloat[Estimated degrees of freedom]{
\begin{tabular}{|rrrrrrr|}  
  % \multicolumn{7}{c}{Estimated degrees of freedom} \\
  \hline
  % Cluster                     &
                                  1                     & 2                     & 3                     & 4                     & 5                     & 6  &7                   \\ \hline
% Complete Case      &
                       24.9 & 9.8 & 11.8 & 12.2 &  39.3&  - & - \\
                       % Full EM     &
                                       % Observed EM    &
24.2 & 9.8 & 14.8 & 161.6 & 200.0 & 9.2 & 26.9 \\ 
 29.7 & 9.8 & 14.7 & 42.1 & 79.0 & 7.8 & 71.3 \\ 
  \hline
\end{tabular}
}\label{tab:grbests}
\end{table}
The three solutions disagree mildly in the estimated mixing
proportions (but ``Observed EM'' tracks ``Full EM'' fairly well).
The degrees of freedom are also different in a few cases. While the {\sc teigen} solution in
\citet{chattopadhyayandmaitra18} found (all five $\nu$s to
be 200), our ``Complete Case'' solution found much fatter-tailed
$t$-mixture components. Figure~\ref{fig:grb} displays the estimated cluster means
and scale parameters of the multivariate $t$-components. We note minor
differences in the means and the scale parameters, with the ``Observed
EM'' and ``Full EM'' cluster parameters being qualitatively the same.
%and visualize the estimated cluster means via parallel coordinates in
%Figure~\ref{fig:grbmu} and cluster dispersions via heat maps in Figure~\ref{fig:grbsig}. 
\begin{comment}
\begin{figure}[h!]
\centering
\caption{Results for the GRB analysis with the BIC-preferred $K = 6$: estimated dispersions $\hat{\bm{\Sigma}}_1, \dots, \hat{\bm{\Sigma}}_K$ for the complete case, full EM, and proposed observed EM approach.}
\label{fig:grbsig}
\end{figure}
\end{comment}
For the seven-groups solution, the ``Full EM'' had slightly higher
log likelihood than the others, however, the main astrophysical properties  
of the solution is similar for both  ``Observed EM'' and ``Full EM''
groups so we only discuss the ``Observed EM'' results.

\citet{mukherjeeetal98} provided a novel approach to describing the
properties of GRBs. This approach, also adopted by
\citet{chattopadhyay17,chattopadhyayandmaitra18,berryandmaitra19,almodovarandmaitra20},
uses the average duration ($T_{90}$), total fluence ($F_ t = F_1 + F_2 + F_3 +
F_4$), and spectral hardness ($H_{ 321} = F_3 /(F_1 + F_2)$) to
characterize the GRBs. (Note that these calculations use the GRB
features in the original scale.) Using these values, we can classify
the seven GRB ``Observed EM'' groups as 
long/bright/soft, short/faint/intermediate, long/faint/soft, short/faint/hard,
long/intermediate/soft, very short/faint/hard, and short/intermediate/soft,
% long/very bright/soft, ultra-long/bright/soft,
% short/faint/intermediate, long/intermediate/soft, short/faint/very
% hard and  short/faint/hard
in terms of their average  duration/fluence/hardness. Further analysis
of our results is  
outside the purview of this paper, but our groups are
able to characterize GRBs  more distinctly compared that in
\citet{chattopadhyay17,chattopadhyayandmaitra18,berryandmaitra19,almodovarandmaitra20}
that used  only the complete cases. 
\section{Discussion} \label{sec:discussion}
In this paper, we consider model-based clustering of partially
recorded or otherwise incomplete data using all and only the observed
values through the use of an observed data model. A corresponding
approximate AECM
algorithm for clustering of partially recorded data is developed and
implemented in the {\tt R} package {\sc MixtClust}. When fitting
finite mixtures of $t$ distributions to incomplete data for the
purpose of clustering, integrating over the missing components has
several benefits compared to complete case analysis or including the
missing components in an EM algorithm: fewer computations are required
in each EM iteration and the approach in our experiments appears to offer somewhat greater resistance to
severe violations of the MCAR assumption. Based on the simulation
experiments of Section~\ref{sec:sim}, we conclude that our approach is
efficient and robust when compared to the corresponding complete case
analysis and full EM based on finite mixture modeling with
multivariate $t$ distributions. We also use our methodology to
characterize  GRBs in the BATSE 4Br catalog into seven
sub-types with distinct and interpretable astrophysical properties.

Further consideration of the relative strengths and
weaknesses between ``Full EM'' and our ``Observed EM'' approach is
warranted. In particular, we need to get a better understanding for
the reason why ``Observed EM'' performs better than ``Full EM''
with increasing deviations from MAR assumptions.
The ``Full EM'' approach utilizes
information on the observed values to inform the missing values, and
we surmise that this information is beneficial for clustering performance
when it is not (too) wrong, in contrast with  the NMAR2 setting where
``Observed EM'' does better. However, as noted by the reviewer,
we also believe that incorporating the missingness mechanism in the
development of the modeling and the inference is a better approach and
should be adopted: however, this is application-specific and requires
knowledge of the specific mechanism causing the missingness. In the
absence of information on missingness, we find that our approach does
well, and is also faster, not only in terms of per iteration, but also
in terms of the time to convergence. Further,
our simulation experiments used the same number of initialization
steps and convergence criteria for all methods, without regard to the
fact that our ``Observed EM'' approach is by design faster than ``Full
EM'', so it would be interesting to compare performance with times set
to be the same, and to see if we can recover the lost ground to ``Full
EM'' using more initializations in the cases where we are currently
out-performed. Also, while using the $t$ distribution for each cluster accommodates outliers, at least
relative to a Gaussian distribution, it assumes that the clusters are
ellipsoidally symmetric about their centers. Such an assumption may be unrealistic
in practice, where clusters could be asymmetric. A natural extension
of our work would incorporate skew-$t$ distributions for such cases
or, alternatively, employ a symmetrizing transformation such as the
Box-Cox transformation considered in finite mixture modeling by
\citet{lo12}. Several other lines of improvement are possible for our
method. First, we only consider general covariance structure
dispersion matrices, but in actuality a simpler structure may be
adequate. Accordingly, future work can incorporate a family of
eigen-decomposed covariance structures \citep{banfieldandraftery93}.
Second, we use a lack-of-progress criterion to assess convergence but
it may be better to use alternative strategies such as Aitken's
acceleration \citep{aitken26} to compute an asymptotic estimate of the
log likelihood as proposed by \citet{bohningetal94}. Finally, while we
use BIC to select the optimal number of clusters, this does not
account for the classification uncertainty in the fitted model as
considered by the integrated completed likelihood (ICL) criterion of
\citet{biernackietal00}. Thus, we see that while we have made some
contributions to the goal of model-based clustering of partial
records, a number of issues remain that merit further attention.
%%%%%%%%%%%%%%%%%%%%%%%%%%%%%%%%%%%%%%%%%%%%%%%%%%%%%%%%%%%%%%%%%%%%%%%%%%%%%%%%%%%%%%%%%%%%

%\backmatter

\section*{Acknowledgments}
%This is acknowledgment text~\cite{Elbaum2002}. Provide text here. This is acknowledgment text. Provide text here. This is acknowledgment text. Provide text here. This is acknowledgment text. Provide text here. This is acknowledgment text. Provide text here. This is acknowledgment text. Provide text here. This is acknowledgment text. Provide text here. This is acknowledgment text. Provide text here. This is acknowledgment text. Provide text here. 
The authors thank the Editor, the Associate Editor and two reviewers
whose careful review of an earlier version of this manuscript greatly
improved its content. Our thanks also to Somak Dutta and Carlos
Llosa-Vite for helpful discussions. 
R. Maitra acknowledges support from the United States Department
of Agriculture (USDA) National Institute of Food and Agriculture
(NIFA) Hatch projects IOW03617 and IOW03717. 
The content of this paper however is solely the responsibility of the 
authors and does not represent the official views of the NIFA or the USDA.

%\subsection*{Author contributions}
%This is an author contribution text. This is an author contribution text. This is an author contribution text. This is an author contribution text. This is an author contribution text. 

\section*{Data Availability}
The GRB dataset used in this application is available at \url{https://github.com/emilygoren/MixtClust}.
\appendix
\section{Derivations for the AECM steps}
\label{appendixa}
We provide here derivations for  $\hat{\bm{\mu}}_k$ in
\eqref{eq:cmmu} and $\hat{\bm{\Sigma}}_k$ in \eqref{eq:cmsig}. 
From equation~(84) of \citet{petersonandpedersen12}, we have
\begin{align*}
0\stackrel{\text{set}}{=}  \frac{\partial}{\partial \bm{\mu}_k}
Q_2\left(\bm{\mu}_k, \bm{\Sigma}_k \given \hat{\bm{\Theta}}\right) 
&= \frac{\partial}{\partial \bm{\mu}_k}
 \sum_{i=1}^n \frac{\hat{z}_{ik}}{2}
\bigg[- \hat{w}_{ik}(\bm{y}_i -
                                                                      \bm{\mu}_k)'\bm{O}_i'(\bm{O}_i\bm{\Sigma}_k\bm{O}_i')^{-1}\bm{O}_i(\bm{y}_i - \bm{\mu}_k)  \bigg] \\ &= 
 \sum_{i=1}^n
                                                                      \hat{z}_{ik}\hat{w}_{ik}\bm{O}_i'(\bm{O}_i\bm{\Sigma}_k\bm{O}_i')^{-1}\bm{O}_i
                                                                      (\bm{y}_i
                                                                      -
                                                                      \bm{\mu}_k).
  \mbox{ This implies }\\
% &\quad\stackrel{\text{set}}{=} 0 \\
 \sum_{i=1}^n \hat{z}_{ik}\hat{w}_{ik}\bm{O}_i'(\bm{O}_i\bm{\Sigma}_k\bm{O}_i')^{-1}\bm{O}_i \bm{y}_i 
 & = \sum_{i=1}^n
  \hat{z}_{ik}\hat{w}_{ik}\bm{O}_i'(\bm{O}_i\bm{\Sigma}_k\bm{O}_i')^{-1}\bm{O}_i
   \bm{\mu}_k, \\
  \quad\mbox{or}\quad
                                                                                                               \hat{\bm{\mu}}_k
  & = \left[\left(\sum_{i=1}^n \hat{z}_{ik}\hat{w}_{ik}
                                                                                                               \bm{O}_i'(\bm{O}_i\bm{\Sigma}_k\bm{O}_i')^{-1}\bm{O}_i\right)\right]^{-1}
                                                                                                               \sum_{i=1}^n \hat{z}_{ik}\hat{w}_{ik}
 \bm{O}_i'(\bm{O}_i\bm{\Sigma}_k\bm{O}_i')^{-1}\bm{O}_i \bm{y}_i .
\end{align*}
%where $\sum_{i=1}^n \hat{z}_{ik}\hat{w}_{ik}\diag(\bm{a}_i)$ is invertible if $\sum_{i=1}^n\hat{z}_{ik}\hat{w}_{ik}I(y_{ij} \text{ is observed}) > 0$ for  $1 \le j \le p$.
To reduce computational complexity, we may use the method-of-moments
update for $\hat\mu$:
$$\hat\mu_k =
  \sum_{i=1}^n \frac{\hat{z}_{ik}}{2}
\bigg[\bm{O}_i'\bm{O}_i\bm{\Sigma}_k\bm{O}_i'\bm{O}_i - \hat{w}_{ik}
       \bm{O}_i'\bm{O}_i(\bm{y}_i - \hat{\bm{\mu}}_k)(\bm{y}_i - \hat{\bm{\mu}}_k)'\bm{O}_i'\bm{O}_i
       \bigg],$$
       where $\sum_{i=1}^n \hat{z}_{ik}\hat{w}_{ik}\diag(\bm{a}_i)$ is
       invertible if $\sum_{i=1}^n\hat{z}_{ik}\hat{w}_{ik}I(y_{ij}
       \text{ is observed}) > 0$ for  $1 \le j \le p$, and then check
       to ensure if the $Q_1$ function increases with this update, or
       revert back to the current value if it does  not.  

For the dispersion updates, we considering each pattern of missingness
and use the selection matrices to expand to dimension $p \times p$,
\begin{align*}
0&\quad\stackrel{\text{set}}{=} \frac{\partial}{\partial \bm{\Sigma}_k}
Q_2\left(\hat{\bm{\mu}}_k, \bm{\Sigma}_k \given \hat{\bm{\Theta}}\right) \\
&\quad= \frac{\partial}{\partial \bm{\Sigma}_k}
\sum_{i=1}^n \frac{\hat{z}_{ik}}{2}
\bigg[- \log\lvert\bm{O}_i\bm{\Sigma}_k\bm{O}_i'\rvert - \hat{w}_{ik}(\bm{y}_i - \hat{\bm{\mu}}_k)'\bm{O}_i'(\bm{O}_i\bm{\Sigma}_k\bm{O}_i')^{-1}\bm{O}_i(\bm{y}_i - \hat{\bm{\mu}}_k)  \bigg] \\
&\quad= \frac{\partial}{\partial \bm{\Sigma}_k}
\sum_{i=1}^n \frac{\hat{z}_{ik}}{2}
\bigg[\log\lvert(\bm{O}_i\bm{\Sigma}_k\bm{O}_i')^{-1}\rvert - \hat{w}_{ik}
                  \trace\left\{
                  \bm{O}_i(\bm{y}_i - \hat{\bm{\mu}}_k)(\bm{y}_i - \hat{\bm{\mu}}_k)'\bm{O}_i' (\bm{O}_i\bm{\Sigma}_k\bm{O}_i')^{-1}
       \right\}\bigg] \\
&\quad=
  \sum_{i=1}^n \frac{\hat{z}_{ik}}{2}
\bigg[\bm{O}_i'\left(\bm{O}_i\bm{\Sigma}_k\bm{O}_i'\right)^{-1}\bm{O}_i - \hat{w}_{ik}
       \bm{O}_i'\left(\bm{O}_i\bm{\Sigma}_k\bm{O}_i'\right)^{-1}\bm{O}_i(\bm{y}_i - \hat{\bm{\mu}}_k)(\bm{y}_i - \hat{\bm{\mu}}_k)'\bm{O}_i'\left(\bm{O}_i\bm{\Sigma}_k\bm{O}_i'\right)^{-1}\bm{O}_i
       \bigg], \\
%       &\quad\stackrel{\text{set}}{=} 0 \\
\end{align*}
\begin{comment}
%       &\implies
       \sum_{i=1}^n \hat{z}_{ik}\bm{O}_i'\bm{O}_i\bm{\Sigma}_k\bm{O}_i'\bm{O}_i 
       = \sum_{i=1}^n \hat{z}_{ik}\hat{w}_{ik}
       \bm{O}_i'\bm{O}_i(\bm{y}_i - \hat{\bm{\mu}}_k)(\bm{y}_i - \hat{\bm{\mu}}_k)'\bm{O}_i'\bm{O}_i \\
       &\implies
       \hat{\bm{\Sigma}}_k 
       = \left( \sum_{i=1}^n \hat{z}_{ik}\bm{a}_i\bm{a}_i'  \right)^{\odot -1}
       \odot \left(\sum_{i=1}^n \hat{z}_{ik}\hat{w}_{ik}
                    \diag(\bm{a}_i)(\bm{y}_i - \hat{\bm{\mu}}_k)(\bm{y}_i - \hat{\bm{\mu}}_k)'\diag(\bm{a}_i) \right).
                    \end{align*}
\end{comment}
which does not provide a closed-form solution. We again propose a method-of-moments estimator for
$\hat\Sigma_k$. Specifically, we propose the update
$$       \hat{\bm{\Sigma}}_k 
       = \left( \sum_{i=1}^n \hat{z}_{ik}\bm{a}_i  \bm{a}_i'  \right)^{\odot -1}
       \odot \left(\sum_{i=1}^n \hat{z}_{ik}\hat{w}_{ik}
                    \diag(\bm{a}_i)(\bm{y}_i - \hat{\bm{\mu}}_k)(\bm{y}_i - \hat{\bm{\mu}}_k)'\diag(\bm{a}_i) \right),$$
 and accept it if $Q_2$ increases, or revert back to the current
 solution.

 \section{Initialization Effect on ``Full EM''}
\label{appendixb}
 The experiments of Section~\ref{expts:comp} demonstrated the good
performance of ``Observed EM'' in terms of speed: the method was also
shown to be competitive 
 \begin{figure}
   \centering
   \mbox{
     \subfloat[Time to convergence of ``Full EM'' using Modified em-EM
     (left lobe of violin)
     and Rnd-EM (right lobe) methods.]{\includegraphics[width=.9\linewidth]{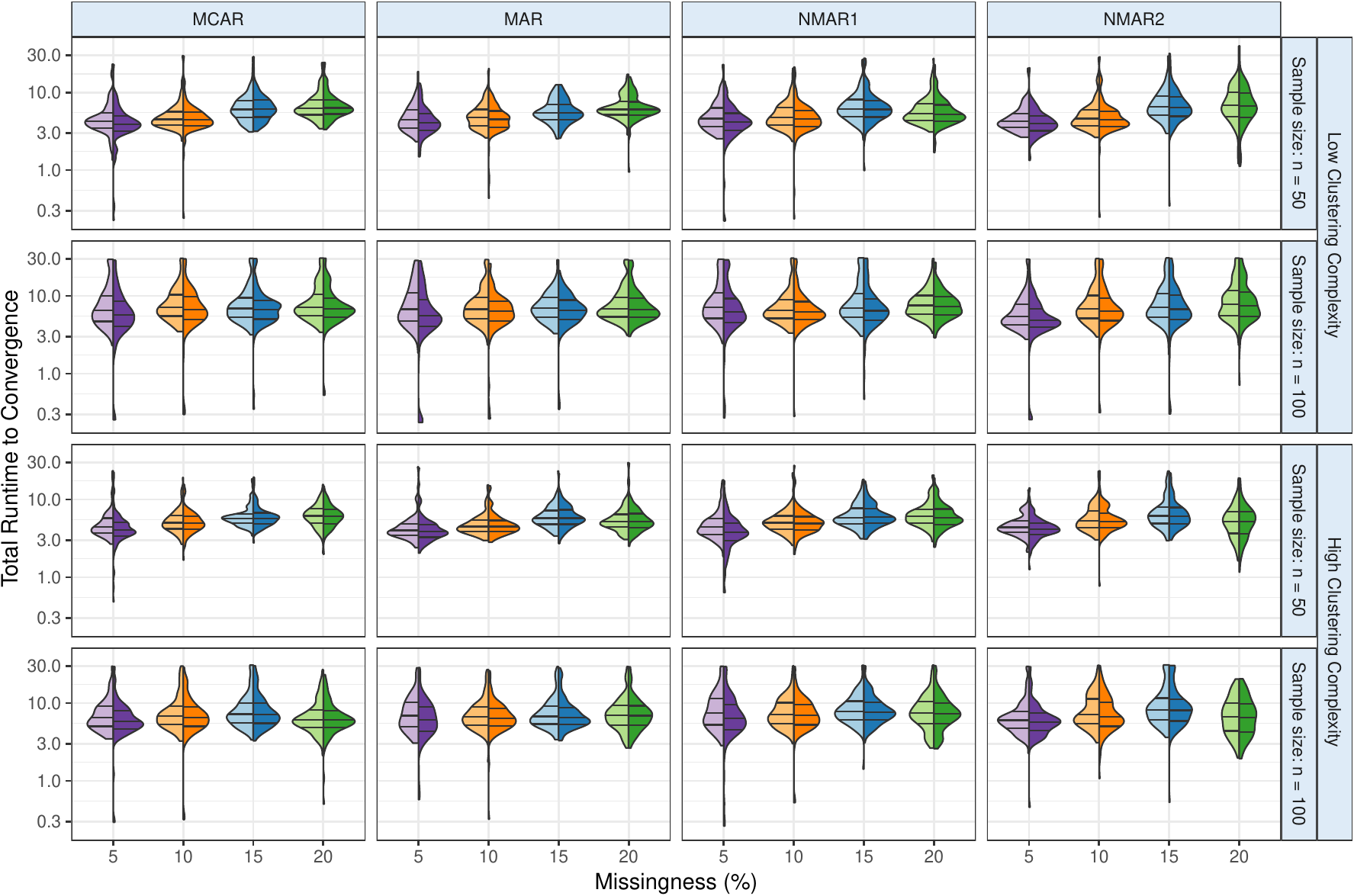}}}
   \mbox{\subfloat[Cluster recovery ability of ``Full EM'' using Modified em-EM
     (left lobe of violin)
     and Rnd-EM (right lobe) methods.]{\includegraphics[width=.9\linewidth]{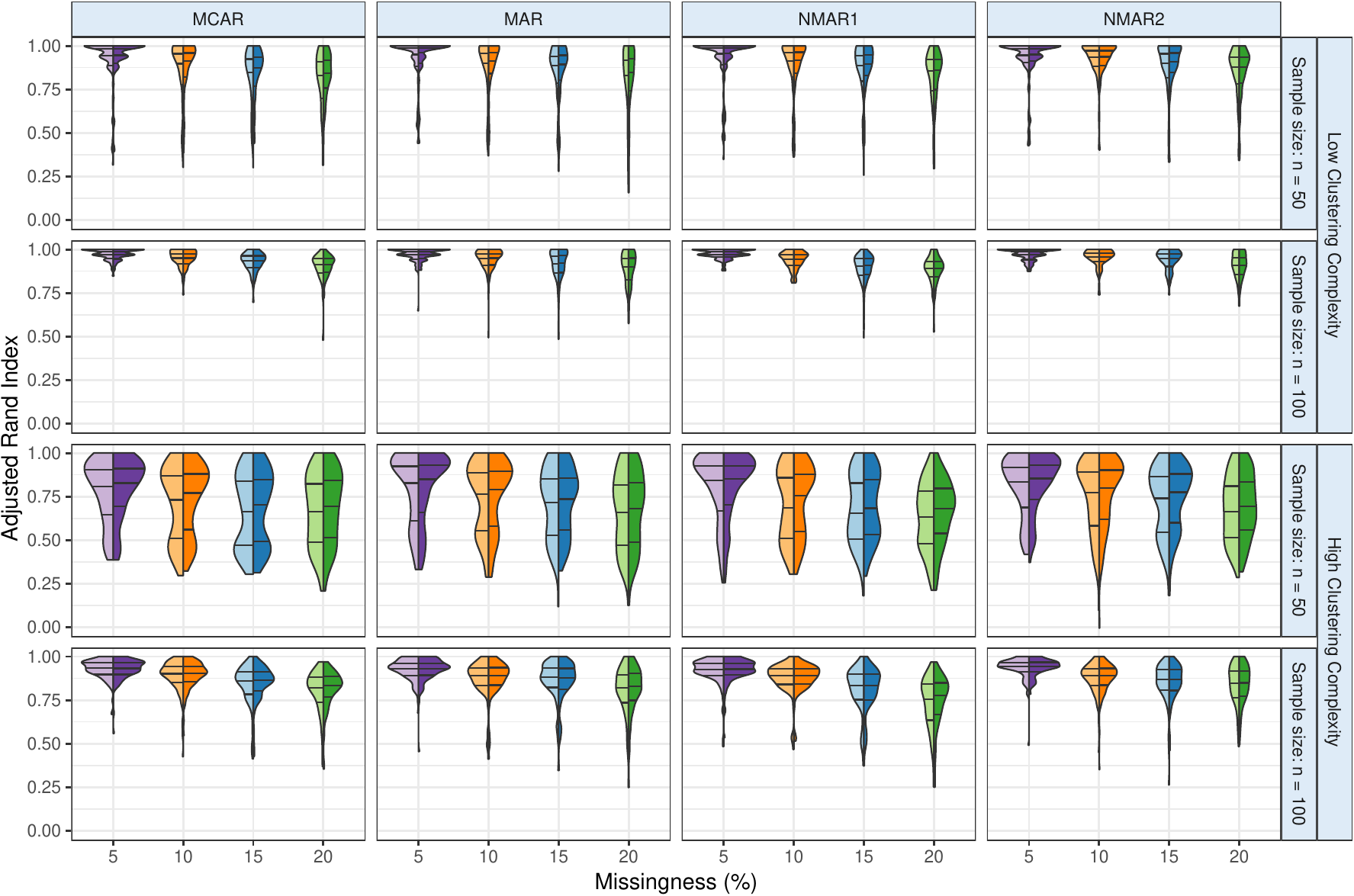}}}
  \caption{Evaluating the (a) speed and (b) performance of the ``Full EM'' algorithm using the 
    Rnd-EM versus Modified em-EM initialization
  methods. Following~\citet{maitra19}, each split violin has
  the Modified em-EM on the left (lighter) lobe and the Rnd-EM on the
  right lobe. The bars denote the three quartiles of each violin.}
%(a) accuracy of BIC in recovering the true  number of clusters
%demonstrated in terms of the difference between the number   of
%clusters   chosen by BIC $\hat{K}$, and the true $K$. (b)   ARI
%values comparing the true partition to the clustering   obtained at $\hat K$.}
% The difference between the number
                                %of clusters   chosen by BIC,
                                %$\hat{K}$, and the true number of
                                %clusters, $K$, is   depicted as a
                                %composition across 100 simulation
                                %replications for    each method,
                                %missingness mechanism, and clustering
                                %complexity   scenario.
%  (b) Adjusted Rand values comparing the true partition to the clustering at the BIC preferred value of $K$ shown as box plots based on 100 simulation replications for each method, missingness mechanism, and clustering complexity scenario.}
\label{fig:sim-fullem}
\end{figure}
in terms of clustering performance. Therefore, we performed another
set of experiments, at the true $K=3$ that used ``Full EM'' with the
Rnd-EM~\citep{maitra09} and Modified em-EM~\citep{maitra13}
evaluations. Both methods used the same number of initializations
($K\sqrt{np}$) and ten long EM runs, but Modified em-EM also used ten
``em'' steps. Figure~\ref{fig:sim-fullem}
displays the results through
a split violin-plot~\citep{maitra19} and indicates that there is not
much difference in the distribution of either the time to convergence or in cluster
recovery ability. We attribute the similar speed of Rnd-EM (with its
zero ``em'' steps) and ``Modified em-EM'' to the fact that the former
ends up often having more iterations in each of the long EM runs.

\bibliography{references}%

\end{document}